\documentclass[journal]{IEEEtran}
\normalsize

\usepackage{amssymb}
\newtheorem{definition}{Definition}

\usepackage{cite}
\ifCLASSINFOpdf
\usepackage[pdftex]{graphicx}
\graphicspath{{fig/}}
\DeclareGraphicsExtensions{.pdf,.jpeg,.png}
\else
\fi

\usepackage[cmex10]{amsmath}
\usepackage{algorithm,algorithmic}
\usepackage{array}
\newcolumntype{C}[1]{>{\centering\let\newline\\\arraybackslash\hspace{0pt}}m{#1}}
\usepackage[tight,footnotesize]{subfigure}

\usepackage{stfloats}
\usepackage{multirow}
\usepackage{booktabs}
\usepackage[flushleft]{threeparttable}
\usepackage{amssymb}
\usepackage{setspace}
\usepackage{stackrel}
\usepackage[pdftex]{graphicx,color}

\newtheorem{remark}{Remark}
\newtheorem{theorem}{Theorem}
\newtheorem{lemma}{Lemma}

\DeclareMathSizes{10}{8}{7}{6}
\IEEEaftertitletext{\vspace{-2\baselineskip}}

\begin{document}
\title{A New Density Evolution  Approximation for LDPC and	Multi-Edge Type LDPC Codes}

\author{Sachini~Jayasooriya,~\IEEEmembership{Student Member,~IEEE,}
	Mahyar~Shirvanimoghaddam,~\IEEEmembership{Member,~IEEE,}    Lawrence~Ong,~\IEEEmembership{Member,~IEEE,}
	Gottfried~Lechner,~\IEEEmembership{Member,~IEEE,}
	and~Sarah~J.~Johnson,~\IEEEmembership{Member,~IEEE}
	\thanks{S. Jayasooriya, M. Shiorvanioghaddam, L. Ong, and S. J. Johnson are with the School of Electrical Engineering and Computer Science, The University of Newcastle, Newcastle, Australia.		G. Lechner is with the Institute for Telecommunications Researchat the University of South Australia, Mawson Lakes, SA 5095, Australia.}}      

\maketitle
%\doublespacing
%\vspace*{-5em}
\begin{abstract}
%\boldmath
This paper considers density evolution for low-density parity-check (LDPC)  and  multi-edge type low-density parity-check (MET-LDPC) codes over the binary input additive white Gaussian noise  channel. We first analyze three single-parameter Gaussian approximations  for density evolution   and discuss  their accuracy  under several conditions, namely at low rates, with punctured and degree-one variable nodes. We  observe that the assumption of symmetric Gaussian distribution  for the  density-evolution messages  is not accurate in the early decoding iterations, particularly at low rates and with punctured variable nodes. Thus single-parameter Gaussian approximation methods produce very poor results in these cases. Based on these observations, we then introduce a new density evolution approximation algorithm for LDPC and MET-LDPC codes. Our method is a combination of full density evolution and a single-parameter Gaussian approximation, where we  assume a  symmetric Gaussian distribution only after density-evolution messages closely follow  a symmetric Gaussian distribution. Our method significantly improves the accuracy of the code threshold estimation. Additionally, the proposed method significantly reduces the computational time of evaluating the code threshold compared to  full density evolution thereby making it more suitable for  code design.
\end{abstract}


\begin{IEEEkeywords}
Belief-propagation,	density evolution,  Gaussian approximation, low-density parity check (LDPC) codes, multi-edge type LPDC codes.
\end{IEEEkeywords}

% \fi
\IEEEpeerreviewmaketitle

%***********************************************************************************************
%***************Section I - Introduction********************************************************
%***********************************************************************************************
\section{Introduction} \label{Introduction}

Graph-based codes, such as low-density parity-check (LDPC), turbo, and repeat-accumulate codes, together with  belief propagation (BP) decoding  have shown to perform extremely close to the Shannon limit with  reasonable decoding complexity~\cite{RichardsonIT2001Design}. These
graph-based codes  can be represented  by  a bipartite Tanner graph in which the variable and check nodes  respectively correspond to the codewords symbols and the parity check constraints~\cite{tannerIT1981recursive}. The error-correcting performance of a code is mainly characterized by the connectivity among the nodes in the Tanner graph where the node degree plays an important role. To specify the node degree distribution in the Tanner graph, the concept of degree distribution in either node perspective or edge perspective is introduced~\cite{lubyIT2001improved}. A code ensemble is then defined as the set of all  codes with a particular degree distribution. As a unifying framework for graph-based codes, Richardson and Urbanke~\cite{RichardsonW2002multi} proposed multi-edge type low density parity-check (MET-LDPC) codes. The benefit of the MET generalization is  greater flexibility in the code structure, which can  improve decoding performances. This generalization is particularly  useful under traditionally difficult requirements, such as high-rate codes with very low error floors or low-rate codes~\cite{RichardsonW2002multi}.

A numerical technique, referred to as Density Evolution (DE), was formulated to analyze the convergence behavior of the BP decoder (i.e., the code threshold) for a given LDPC~\cite{lubyIT2001improved} or MET-LDPC~\cite{richardsonBOOK2008modern} code ensemble under BP decoding,  where the  code threshold is defined as  the maximum channel noise level at which the decoding error probability  converges to zero as the code length goes to infinity. DE determines the performance of BP decoding for a given code ensemble by tracking the probability density function (PDF) of messages passed along the edges in the corresponding Tanner graph through the iterative decoding process. Then, it is possible to  test whether, for a given channel condition and a given degree distribution, the decoder can successfully decode the transmitted message (with the decoding error probability tends to zero as the iterations progress). This allows us to design and optimize LDPC and MET-LDPC  degree distributions using the DE threshold (i.e., the code threshold found using DE) as the cost function.

Calculating thresholds and designing  LDPC and MET-LDPC degree distributions using DE are computationally intensive  as they require numerical evaluations of the PDFs of the messages passed along the Tanner graph edges in each decoding iteration. Because of this, for LDPC codes on the binary input additive white Gaussian noise (BI-AWGN) channel, Chung~\emph{et~al.}~\cite{chungIT2001analysis, chungTHESIS2000} and Lehmann and Maggio~\cite{lehmannIT2003analysis} approximated the message PDFs by Gaussian PDFs, each using a single parameter, to  simplify the analysis.  Existing work concerning Gaussian approximations has relied on four different parameters in order to obtain a single-parameter model of the message PDFs, including mean value of the PDF~\cite{chungIT2001analysis}, bit-error rate (BER)~\cite{lehmannIT2003analysis},  reciprocal-channel approximation (RCA)~\cite{chungTHESIS2000} and  mutual information (e.g., EXIT charts)~\cite{tenCOM2001convergence}. Several papers~\cite{fuICC2006gaussian,ardakaniCOM2004more,chungIT2001analysis,xieISIT2006accuracy} have investigated the accuracy of the Gaussian approximation for BP decoding of standard LDPC codes and shown that it is  accurate for medium-to-high rates.   However in most of the literature regarding DE for MET-LDPC codes, only the full density evolution (full-DE) has been studied~\cite{richardsonBOOK2008modern}.  In full-DE, the quantized PDFs of the messages are passed along the edges without any approximation. Typically, for full-DE, thousands of points are used to accurately describe one message PDF. Schmalen and Brink~\cite{schmalenSCC2013combining} have used the Gaussian approximation based on the mean of the message PDF~\cite{chungIT2001analysis} to  evaluate the behavior of protograph based LDPC codes, which is a subset of MET-LDPC codes.

The contributions of this paper are as follows: 1) We investigate the accuracy of Gaussian approximations  for  BP decoding. We follow the approximation techniques suggested for  LDPC codes~\cite{chungIT2001analysis,chungTHESIS2000,lehmannIT2003analysis}, which describe each  DE-message PDF  using a single parameter. Based on our observations of the evolution of PDFs in the MET-LDPC codes, we found that those Gaussian approximations are not accurate for the scenarios where MET-LDPC codes are useful, i.e., at low rate and with punctured variable nodes. 2) In light of this, we  propose a hybrid-DE method, which combines the  full-DE  and a Gaussian approximation. Our proposed hybrid-DE  allows us to evaluate the code threshold (i.e., the cost function in the  code optimization)  of  MET-LDPC and LDPC code ensembles significantly faster than the   full-DE and with accuracy better than Gaussian approximations.  3) We design good MET-LDPC codes using the  proposed hybrid-DE and  show that the hybrid-DE well approximates the  full-DE for code design.

This paper is organized as follows. Section~\ref{Background} briefly reviews the basic concepts of  MET-LDPC codes. In Section~\ref{Gaussian app}
we extend   Gaussian approximations for LDPC codes to MET-LDPC codes, and in Section~\ref{validity of GA}, we discuss the accuracy of the Gaussian approximations under the conditions where MET-LDPC codes are  more beneficial. Section~\ref{Hybrid DE} presents the proposed hybrid-DE method, and  Section~\ref{code design GA}  demonstrates the benefits of code design using the proposed hybrid-DE method over existing Gaussian approximations. Finally, Section~\ref{Conclusion} concludes the paper.

%***********************************************************************************************
%***************Section II - Background*********************************************************
%***********************************************************************************************
\section{Background of MET-LDPC codes} \label{Background}

\subsection{MET-LDPC code ensemble}

Unlike standard LDPC code ensembles where the graph connectivity is constrained only by the node degrees, in the multi-edge setting, several edge-types can be defined, and every node is characterized by the number of connections to edges of each edge-type. Within this framework, the degree distribution of MET-LDPC code ensemble can be specified through two node-perspective multinomials related to  the variable nodes and check nodes respectively~{\protect\cite[page 383]{richardsonBOOK2008modern}}: \vspace*{-1em}
\begin{align}
L(\boldsymbol{r},\boldsymbol{x}) &= \sum L_{\boldsymbol{b},\boldsymbol{d}}~ \boldsymbol{r}^{\boldsymbol{b}}~ \boldsymbol{x}^{\boldsymbol{d}}			\label{eq : MET_LDPC_lamda}\\
R(\boldsymbol{x}) &= \sum R_{\boldsymbol{d}} ~\boldsymbol{x}^{\boldsymbol{d}},			
\label{eq : MET_LDPC_roh}
\end{align}
where $\boldsymbol{b}, \boldsymbol{d}, \boldsymbol{r}$ and $\boldsymbol{x}$ are vectors defined as follows. Let $m_e$ denote the number of edge-types corresponding to the graph  and  $m_r$ denote the number of different channels over which codeword bits may be transmitted. A vector $\boldsymbol{d} = [d_1, \dots, d_{m_e}]$ is defined for each node in the graph, where $d_i$ is the number of edges of the $i{\text{th}}$ edge-type connected to that node, and  we use $\boldsymbol{x}^{\boldsymbol{d}}$ to denote $\prod_{i=1}^{m_e} x_i^{d_i}$.  As the variable nodes  receive information from the channel over which the codeword bits are transmitted, there is an additional vector $\boldsymbol{r}^{\boldsymbol{b}}=\prod_{i=0}^{m_r} r_i^{b_i}$ associated with each variable  node where $b_i$ designates the type of the message (i.e., the message PDF) it receives from the channel. Typically, $\boldsymbol{b} = [b_0,\dots, b_{m_r}]$ has only two entries since in a BI-AWGN channel, a codeword bit is either    punctured  (the codeword bits not transmitted: $\boldsymbol{b} = [1,0]$ ) or  transmitted through a single channel ($\boldsymbol{b} = [0,1]$). Finally, $L_{\boldsymbol{b},\boldsymbol{d}}$ and $R_{\boldsymbol{d}}$ are non-negative real numbers corresponding to the fraction of variable nodes of type ($\boldsymbol{b}, \boldsymbol{d}$) and the fraction of check nodes of type $\boldsymbol{d}$ in the ensemble, respectively.

%figure for example MET-LDPC tanner graph
\begin{figure}[t!]
	\centering
	\includegraphics[width=0.5\linewidth]{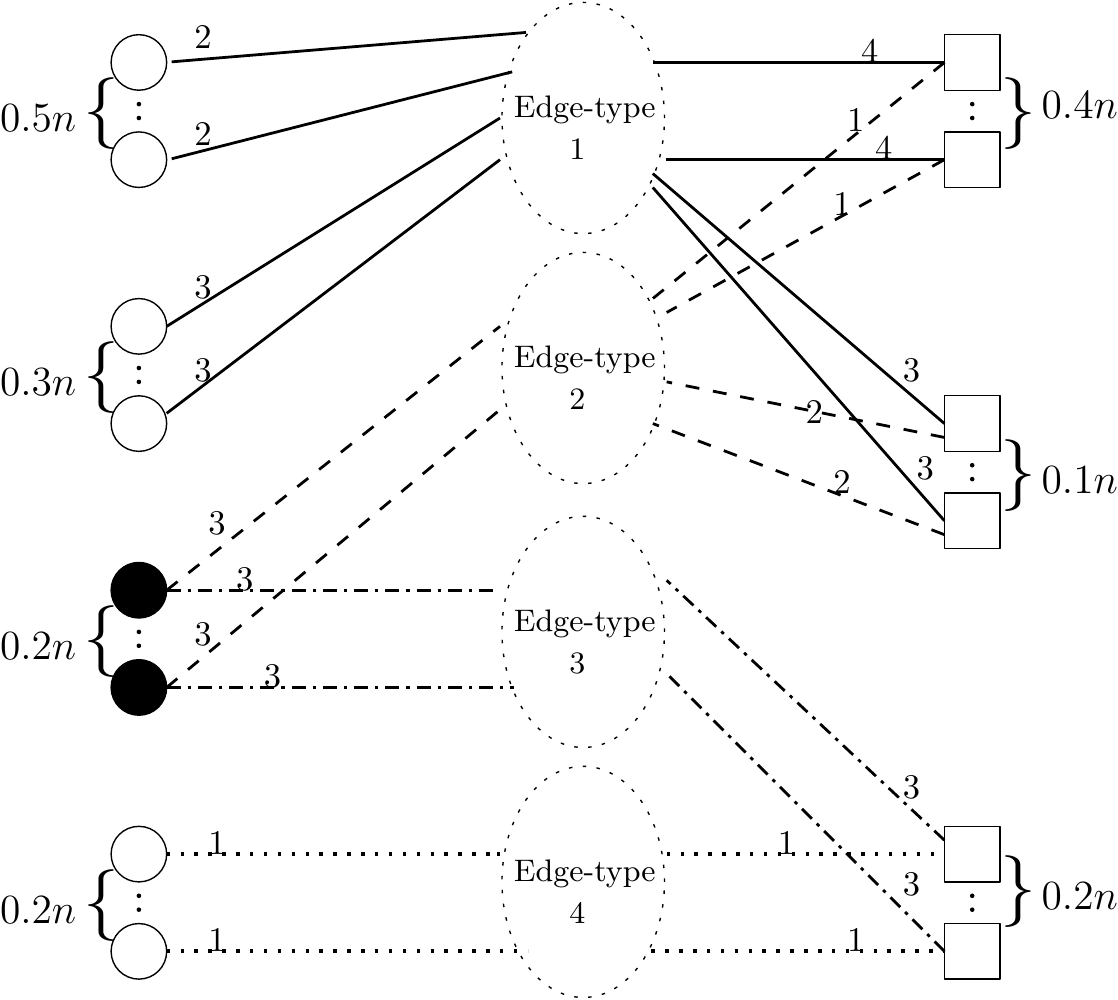}
	\vspace*{-0.5em}
	\caption{Graphical representation of a four-edge type MET-LDPC code (Table VI in~\cite{RichardsonW2002multi}), where \textquoteleft $\circ$\textquoteright~  (resp., \textquoteleft $\bullet$\textquoteright ) represents the unpunctured (resp., punctured) variable nodes and \textquoteleft $\Box$\textquoteright~ represents the check nodes. The number of nodes for different edge-types are shown as fractions of the code length $n$, where $n$ is the number of transmitted (i.e. unpunctured) codeword bits.}
	\label{fig:MET_Tanner}
	%\vspace*{-2em}
\end{figure}

Node-perspective degree distributions can be converted to edge-perspective via  the following multinomials, where $\lambda_i(\boldsymbol{r},\boldsymbol{x})$ and $\rho_i(\boldsymbol{x})$ are related to  the variable nodes and check nodes, respectively~{\protect\cite[pages 390-391]{richardsonBOOK2008modern}}: 
\begin{multline}
\label{lamda_MET}
\big(\lambda_1(\boldsymbol{r},\boldsymbol{x}),\lambda_2(\boldsymbol{r},\boldsymbol{x}),\dots,\lambda_{m_e}(\boldsymbol{r},\boldsymbol{x})\big) = \\ \left( \frac{L_{x_1}(\boldsymbol{r},\boldsymbol{x})}{L_{x_1}(\boldsymbol{1},\boldsymbol{1})}, \frac{L_{x_2}(\boldsymbol{r},\boldsymbol{x})}{L_{x_2}(\boldsymbol{1},\boldsymbol{1})},\dots, \frac{L_{x_{m_e}}(\boldsymbol{r},\boldsymbol{x})}{L_{x_{m_e}}(\boldsymbol{1},\boldsymbol{1})} \right)
\end{multline}
\begin{multline}
	\label{rho_MET}
	\big(\rho_1(\boldsymbol{x}),\rho_2(\boldsymbol{x}),\dots,\rho_{m_e}(\boldsymbol{x})\big) = \\ \left( \frac{R_{x_1}(\boldsymbol{x})}{R_{x_1}(\boldsymbol{1})}, \frac{R_{x_2}(\boldsymbol{x})}{R_{x_2}(\boldsymbol{1})},\dots, \frac{R_{x_{m_e}}(\boldsymbol{x})}{R_{x_{m_e}}(\boldsymbol{1})} \right),
\end{multline}
where  	 	
\begin{align*}
L_{x_i}(\boldsymbol{r},\boldsymbol{x}) &= \frac{\partial}{\partial{x_i}} L(\boldsymbol{r},\boldsymbol{x}) \\
L_{x_i}(\boldsymbol{1},\boldsymbol{1}) &= \frac{\partial}{\partial{x_i}} L(\boldsymbol{r},\boldsymbol{x})\biggr|_{\boldsymbol{r}=\boldsymbol{1},\boldsymbol{x}=\boldsymbol{1}} \\
R_{x_i}(\boldsymbol{x}) &= \frac{\partial}{\partial{x_i}} R(\boldsymbol{x})\\
R_{x_i}(\boldsymbol{1}) &= \frac{\partial}{\partial{x_i}} R(\boldsymbol{x})\biggr|_{\boldsymbol{x}=\boldsymbol{1}}.
\end{align*}
The  rate of a MET-LDPC code is given by 
\begin{align}
	r &= L(\boldsymbol{1},\boldsymbol{1}) - R(\boldsymbol{1}),
	\label{code rate}
\end{align}
where $\boldsymbol{1}$ denotes a vector of all $1$'s with the length determined by the context.

A rate $1/2$  MET-LDPC code ensemble is shown in Fig.~\ref{fig:MET_Tanner}, where the node-perspective degree distributions are given by $L(\boldsymbol{r},\boldsymbol{x}) = 0.5r_1x_1^2+0.3r_1x_1^3+0.2r_0x_2^3x_3^3+0.2r_1x_4$ and $R(\boldsymbol{x}) = 0.4x_1^4x_2+0.1x_1^3x_2^2+0.2x_3^3x_4$. Here $r_0$ denotes punctured nodes and $r_1$ denotes unpunctured nodes.

\subsection{BP decoding and density evolution for MET-LDPC codes}
In the BP decoding algorithm,  messages are passed along the edges of the Tanner graph from variable nodes to their incident check nodes and vice versa until a valid codeword is found or  a predefined maximum number of decoding iterations has been reached. Each BP decoding iteration involves two steps.
\begin{enumerate}
	\item {A variable node processes the  messages it receives from its neighboring check nodes and from its corresponding channel and outputs messages  to its neighboring check nodes.}
	\item{A check node processes  inputs from its neighboring variable nodes and passes messages back to its neighboring variable nodes.}
\end{enumerate}
In most cases of binary codes\footnotemark[1]\footnotetext[1]{Throughout this paper, we assume that the all-zero codeword is sent. } transmitted on BI-AWGN channel, these BP decoding messages\footnotemark[2]\footnotetext[2]{The BP decoding messages received by every node at every iteration are independent and identically distributed.} are  expressed as  log-likelihood ratios (LLRs)~{\protect\cite[pages 213-226]{ryanBook2009channel}}. This is used to reduce the complexity of the BP decoder, as multiplication of message probabilities corresponds to the summation of  corresponding LLRs.

Let $v_e^{(\ell)}$ denote the  message LLR  sent by a variable node to a check node along edge $e$, (i.e, variable-to-check) at the $\ell{\text{th}}$ iteration of the BP decoding, and $u_e^{(\ell)}$ denote the message LLR  sent by a check node to a variable node along edge $e$, (i,e, check-to-variable) at the $\ell{\text{th}}$ iteration of the BP decoding.  At the variable node, $v_e^{(\ell)}$  is computed based on the observed channel LLR ($u_0$) and message LLRs received from the neighboring check nodes except for the incoming message on the current edge for which the output message is being calculated. Thus the variable-to-check  message on edge $e$ at the $\ell$th decoding iteration is as follows: %~\cite{RichardsonIT2001capacity}:
\begin{align}
v_e^{(\ell)} =
\begin{cases}
u_0 & \text{if~} \ell=1,\\
u_0 + \sum_{i\neq e}u_i^{(\ell-1)} & \text{if~} \ell>1.
\end{cases}
\label{eq.VN update}
\end{align}
The message outputs on edge $e$ of a check node at the $\ell$th decoding iteration can be obtained from the  ``tanh rule'': %~\cite{RichardsonIT2001capacity}: \vspace*{-1em}
\begin{align}
u_e^{(\ell)} &= 2 \tanh^{-1} \left(\prod_{j\neq e}\tanh\left(\frac{v_j^{(\ell)}}{2}\right) \right).
\label{eq.CN update}
\end{align}
For more details we refer readers to Ryan and Lin~{\protect\cite[pages 201-248]{ryanBook2009channel}}.

DE is the main tool for  analyzing  the  average asymptotic behavior of the BP decoder for MET-LDPC code ensembles, when  the block length goes to infinity. For  BP decoding on a BI-AWGN channel, these LLR values (i.e., $v, u,u_0$) are continuous random variables, thus can be described by PDFs for analysis using DE. To analyze the evolution of theses PDFs in the BP decoder, we define $f(v^{(\ell)}), f(u^{(\ell)}), f(u_0)$ which denote the PDF of the variable-to-check message, check-to-variable message and channel LLR, respectively. Unlike standard LDPC codes, in the MET framework, because of the existence of multiple edge-types, only the incoming messages from same edge-type are assumed to be  identically distributed. However, all the incoming messages are assumed to be  mutually independent. Recall that in MET-LDPC codes, a variable node is identified by its type, ($\boldsymbol{b},\boldsymbol{d}$), and a check node by its type, $\boldsymbol{d}$.  Thus from~(\ref{eq.VN update}) the PDF of the variable-to-check message from a variable node  type, ($\boldsymbol{b},\boldsymbol{d}$), along edge-type $i$  at the $\ell{\text{th}}$ decoding iteration can be written as follows~{\protect\cite[pages 390-391]{richardsonBOOK2008modern}}:
\begin{multline}
\label{eq:VN to CN PDF MET}
f\big({v_{\boldsymbol{b},\boldsymbol{d}}}^{(\ell)}(i)\big) = f\big(u_{\boldsymbol{b}}\big) \bigotimes \left[f\big({u}^{(\ell-1)}(i)\big)\right]^{\bigotimes (d_i-1)} \\ \bigotimes_{k=1, k\neq i}^{m_e}\left[ f\big({u}^{(\ell-1)}(k)\big)\right]^{\bigotimes d_k},
\end{multline}
where  $\bigotimes$ denotes convolution. The $d_i$-fold and $(d_i-1)$-fold convolutions follow from the assumption that the incoming messages from a check node along edge-type $i$ are independent and identically distributed and  $f\big({u}^{(\ell-1)}(i)\big)$ is used to denote this common PDF.
$f\big(u_{\boldsymbol{b}}\big)$ is the PDF of the channel LLR.

From~(\ref{eq.CN update}) the PDF of the check-to-variable message from a check node  type, $\boldsymbol{d}$, along edge-type $i$  at the $\ell{\text{th}}$ decoding iteration can be calculated as follows~{\protect\cite[pages 390-391]{richardsonBOOK2008modern}}:
\begin{align}
\label{eq:CN to VN PDF MET}
f\big({u_{\boldsymbol{d}}}^{(\ell)}(i)\big) =  \left[f\big({v}^{(\ell)}(i)\big)\right]^{\boxtimes (d_i-1)}  \stackrel[k=1, k\neq i]{m_e}{\boxtimes} \left[ f\big({v}^{(\ell)}(k)\big)\right]^{\boxtimes d_k}, 
\end{align}
where  $f\big({v}^{(\ell)}(i)\big)$ is  the PDF of the message from a variable node along edge-type $i$ at the $\ell{\text{th}}$ decoding iteration.  The computation of $f\big(u^{(\ell)}\big)$ is not as straightforward as that for $f\big(v^{(\ell)}\big)$ and requires the transformation of $\log(.)$ and $\log^{-1}(.)$. So we use $\boxtimes$ to denote the  convolution  when computing the PDF of  $f\big(u^{(\ell)}\big)$ for check-to-variable  messages. For more details we refer readers to Richardson and Urbanke~{\protect\cite[pages 390-391, 459-478]{richardsonBOOK2008modern}}.

%***********************************************************************************************
%***************Section III - Gaussian approximations to density evolution**********************
%***********************************************************************************************

\section{Gaussian approximations to density evolution for MET-LDPC codes} \label{Gaussian app}
In this section, we consider MET-LDPC codes  over BI-AWGN channels with Gaussian approximations to DE. As already shown  for LDPC codes~\cite{chungIT2001analysis,ardakaniCOM2004more}, the PDFs of variable-to-check and check-to-variable messages can be close to a Gaussian distribution in certain cases, such as when check node degrees are small and variable node degrees are relatively large. Since a Gaussian PDF can be completely specified by its mean ($m$) and variance ($\sigma^2$), we need to track only these two parameters during the BP decoding algorithm. Furthermore, it was shown by Richardson~\emph{et~al.}~\cite{RichardsonIT2001Design} that the PDFs of variable-to-check and check-to-variable messages and channel inputs  satisfy the symmetry condition: $f(x) = \mathrm{e}^xf(-x)$ where $f(x)$ is the PDF of variable $x$. This condition greatly simplifies the analysis because it implies  $\sigma^2=2m$ and reduces the description of the PDF to a single parameter.  Thus, by tracking the changes of the mean ($m$) during iterations, we can determine the evolution of the PDF of the check node message, $f(u^{(\ell)}) = \mathcal{N}(m_u^{(\ell)},2m_u^{(\ell)})$ and the variable node message, $f(v^{(\ell)}) = \mathcal{N}(m_v^{(\ell)},2m_v^{(\ell)})$
where $\mathcal{N}(m,\sigma^2)$ is the Gaussian PDF with mean  $m$ and variance $\sigma^2$. $m_v^{(\ell)}$ and $m_u^{(\ell)}$ are the mean of the  variable-to-check and check-to-variable messages, respectively.

\subsection{Approximation 1: Gaussian approximation based on the mean of the message PDF}\label{Approximation 1}
In this subsection, we will extend   the Gaussian approximation method proposed by Chung~\emph{et~al.} ~\cite{chungIT2001analysis}  for the threshold estimation  of standard LDPC codes  to that of MET-LDPC codes.  This  method is based on approximating message PDFs  using a single parameter, i.e., the mean of the message PDF.

Recall that a variable node is identified by its  type, ($\boldsymbol{b},\boldsymbol{d}$), and a check node by its  type, $\boldsymbol{d}$. Since the  PDFs of the messages sent by the variable node are approximated as  Gaussian,  the mean of the  variable-to-check message from a variable node  type, ($\boldsymbol{b},\boldsymbol{d}$), along edge-type $i$  at the $\ell{\text{th}}$ decoding iteration is given by 
\begin{align}
\label{eq:VN to CN MET maen}
m_{v_{\boldsymbol{b},\boldsymbol{d}}}^{(\ell)}(i) = m_{u_{\boldsymbol{b}}} + (d_i-1)m_{u}^{(\ell-1)}(i) + \sum_{k=1, k\neq i}^{m_e} d_k  m_{u}^{(\ell-1)}(k),
\end{align}
\noindent where  $m_{u_{\boldsymbol{b}}}$ is the mean of the message from the channel and $m_{u}^{(\ell-1)}(i)$ is the mean  of the check-to-variable message along edge-type $i$ at the $(\ell-1){\text{th}}$ decoding iteration. The updated mean of   the check-to-variable  message from check node type of $\boldsymbol{d}$ along edge-type $i$  at the $\ell{\text{th}}$ decoding iteration can be written as 
\begin{multline}
\label{eq:mu_b}
m_{u_{\boldsymbol{d}}}^{(\ell)}(i) =   \phi^{-1}\big( 1- \left[ 1- \phi(m_v^{(\ell)}(i))\right]^{d_i-1}  \\ \prod_{k=1,k\neq i}^{m_e}  \left[ 1- \phi(m_v^{(\ell)}(k))\right]^{d_k} \big),
\end{multline}
\noindent where $m_{v}^{(\ell)}(i)$ is the mean of  the variable-to-check message along edge-type $i$ at the $\ell{\text{th}}$ decoding iteration. The  mean of  the variable-to-check  and the check-to-variable messages along edge-type $i$ at the $\ell{\text{th}}$ decoding iteration is given by 
\begin{align}
m_{v}^{(\ell)}(i) &= \sum_{\boldsymbol{d}} \lambda_{i_{\boldsymbol{d}}} m_{v_{\boldsymbol{b},\boldsymbol{d}}}^{(\ell)}(i) \label{avg_mv}\\
m_{u}^{(\ell)}(i) &= \sum_{\boldsymbol{d}} \rho_{i_{\boldsymbol{d}}} m_{u_{\boldsymbol{d}}}^{(\ell)}(i), \label{avg_mu}
\end{align}
where $\lambda_i$ and $\rho_i$ are the variable and  check node edge-degree distributions with respect to edge-type $i$, respectively and 
\begin{align*}
\phi(x) =
\begin{cases}
1- \frac{1}{\sqrt{4\pi x}}\int_{\mathbb{R}} \tanh(\frac{u}{2}) \mathrm{e} ^{-(u-x)^2 / (4x)} d{u},  & \text{if } x>0\\
1,  & \text{otherwise.}
\end{cases}
\end{align*}
It is important to note that $\phi(x)$ is continuous and monotonically decreasing over $[0,\infty)$ with $\phi(0)=1$ and $\phi(\infty)=0$~\cite{chungIT2001analysis}.
	
\subsection{Approximation 2: Gaussian approximation based on the bit error rate} \label{Approximation 2}
In this subsection, we will extend a Gaussian approximation method proposed by Lehmann~\emph{et~al.}~\cite{lehmannIT2003analysis} that estimates thresholds of standard LDPC codes to that of MET-LDPC codes.  This  method is based on a closed-form expression in terms of error probabilities (i.e., the probability that a variable node is sending an incorrect message).

Consider a check node of type, $\boldsymbol{d}$. The  error probability  of a check-to-variable  message  from a check node type, $\boldsymbol{d}$  along edge-type $i$ at the $\ell{\text{th}}$ decoding iteration is given by 
\begin{multline}
\label{eq:MET_ber1}
P_{u_{\boldsymbol{d}}}^{(\ell)}(i) = \frac{1}{2} \big[1-  \big( 1-2P_{v}^{(\ell-1)}(i)\big)^{d_i - 1} \\
\prod_{k=1,k\neq i}^{m_e}\big( 1-2P_{v}^{(\ell-1)}(k)\big)^{d_k} \big],
\end{multline}
\noindent where $P_{v}^{(\ell-1)}(i)$ is the average  error probability  of the variable-to-check  message along edge-type $i$ at the $(\ell-1){\text{th}}$ decoding iteration. Since we  suppose that the all-zero codeword is sent, the error probability of a variable node at the $\ell{\text{th}}$ decoding iteration is simply the  average probability that the variable-to-check messages are negative. We also assume that the PDF of variable-to-check message is symmetric Gaussian; therefore the  error probability of a variable-to-check  message from a variable node type, $(\boldsymbol{b},\boldsymbol{d})$ along edge-type $i$ at the $\ell{\text{th}}$ decoding iteration is given by 
\begin{align}
\label{eq:MET_berVN}
P_{v_{\boldsymbol{b},\boldsymbol{d}}}^{(\ell)}(i)	 &=   Q\left(\sqrt{\frac{m_{v_{\boldsymbol{b},\boldsymbol{d}}}^{(\ell)}(i)}{2}}\right),	
\end{align}
where $m_{v_{\boldsymbol{b},\boldsymbol{d}}}^{(\ell)}$ is the mean of the  variable-to-check message from a variable node  type, ($\boldsymbol{b},\boldsymbol{d}$), along edge-type $i$  at the $\ell{\text{th}}$ decoding iteration,  and 
\begin{align}
Q(x)	 &=  \frac{1}{\sqrt{2\pi}}\int_x^{+\infty} \mathrm{e}^{\frac{-t^2}{2}} d{t}.
\end{align}
$m_{v_{\boldsymbol{b},\boldsymbol{d}}}^{(\ell)}$ can be calculated using (\ref{eq:VN to CN MET maen}) by substituting  
\begin{align}
m_u^{(\ell)}(i) &= 2 \left( Q^{-1} (P_u^{(\ell)}(i)\right)^2
\end{align}
for each $m_u^{(\ell)}(i)$, where $P_{u}^{(\ell)}(i)$ is the average  error probability  of the check-to-variable  message along edge-type $i$ at the $\ell{\text{th}}$ decoding iteration. The  average error probability  of  the variable-to-check  and the check-to-variable messages along edge-type $i$ at the $\ell{\text{th}}$ decoding iteration is given by 
\begin{align}
P_{v}^{(\ell)}(i) &= \sum_{\boldsymbol{d}} \lambda_{i_{\boldsymbol{d}}} P_{v_{\boldsymbol{b},\boldsymbol{d}}}^{(\ell)}(i)\\
P_{u}^{(\ell)}(i) &= \sum_{\boldsymbol{d}} \rho_{i_{\boldsymbol{d}}} P_{u_{\boldsymbol{d}}}^{(\ell)}(i),
\end{align}
where $\lambda_i$ and $\rho_i$ are the variable and  check node edge-degree distributions with respect to edge-type $i$, respectively.

\subsection{Approximation 3:   Gaussian approximation based on the reciprocal-channel approximation }\label{Approximation 3}
In this subsection, we will extend another Gaussian approximation method, proposed by Chung~{\protect\cite[pages 189-193]{chungTHESIS2000}} to estimates thresholds of regular LDPC codes, to that of MET-LDPC codes.  This  method is called reciprocal-channel approximation (RCA),  which is based on reciprocal-channel mapping and  mean ($m$) of the  node message is used  as the one-dimensional tracking parameter for the BI-AWGN channel.

With the RCA technique in DE, $m$ is additive at the variable nodes similar to Approximation 1 (see (\ref{eq:VN to CN MET maen})). The difference between Approximation 1 and Approximation 3 is how the check nodes calculate their output messages. Instead of evaluating  the $\tanh$ function in Approximation 1, Approximation 3 uses the reciprocal-channel mapping, $\psi(m)$, which is additive at the check nodes.  $\psi(m)$ is defined as follows~{\protect\cite[pages 189-193]{chungTHESIS2000}}:
\begin{align}
	\psi(m) &= C^{-1}_{\text{AWGN}}(1-C_{\text{AWGN}}(m)),
\end{align}
where $C_{\text{AWGN}}(m)$ is the capacity  of the BI-AWGN channel  as a function of the mean of the channel message, and 
\begin{align*}
	C_{\text{AWGN}}(m) &=1 - \frac{1}{2\sqrt{\pi m}}  \int_{-\infty}^{\infty} \log_2 (1+\mathrm{e}^{-x}) \mathrm{e}^{\frac{-(x-m)^2}{4m}} dx,
\end{align*}
Then the mean of  check-to-variable message from a check node type, $\boldsymbol{d}$  along edge-type $i$  at the $\ell{\text{th}}$ decoding iteration is given by 		
\begin{align}
	\psi(m_{u_{\boldsymbol{d}}}(i)^{(\ell)}) =  (d_i-1)\psi(m_v^{(\ell)}(i)) + \sum_{k=1, k\neq i}^{m_e} d_k  \psi(m_v^{(\ell)}(k)).
\end{align}
$m_{v}^{(\ell)}(i)$ and $m_{u}^{(\ell)}(i)$ can be calculated from~(\ref{avg_mv}) and~(\ref{avg_mu}), respectively.

%***********************************************************************************************
%***************Section IV - Validity of Gaussian approximations to density evolution***********
%***********************************************************************************************

\section{Validity of the Gaussian assumption for density evolution  }\label{validity of GA}
As we discussed in Section~\ref{Background}, in the BP decoder, there are three types of messages: the channel message, the variable-to-check message, and the check-to-variable message. We analyze the PDF of these messages on the BI-AWGN channel to  evaluate the Gaussian assumption for DE message PDFs.

\subsection{Channel messages}
Let $c=(c_1, c_2,\dots)$ be a binary codeword ($c_i\in \{0,1\}$) on a BI-AWGN channel. A codeword bit, $c_i$ can be mapped to the transmitted symbol $x_i=1$ if $c_i=0$ and $x_i=-1$ otherwise. Then, the $i$th received symbol at the output of the AWGN channel is $y_i = \sqrt{E_c}x_i + z_i$ where $E_c$ is the energy per transmitted symbol and $z_i$ is the AWGN, $z_i \sim \mathcal{N} (0,\sigma_n^2)$.  The  LLR ($\mathcal{L}(\cdot)$) for the received signal, $y_i$ is given by  
\begin{align*}
u_0 = \mathcal{L}(x_i|y_i)  &= \log\frac{\Pr(y_i | x_i=1)}{\Pr(y_i | x_i=-1)}
= \frac{2\sqrt{E_c}}{\sigma_n^2}y_i.
\end{align*}
Assuming that the all-zero codeword is sent and that $\sqrt{E_c}$ is 1, \vspace*{-1em}
\begin{align}
u_0 = \mathcal{L}(x_i|y_i) &= \frac{2}{\sigma_n^2}y_i,
\end{align}
which is a Gaussian random variable with
$\text{E}[{u_0}] = \frac{2}{\sigma_n^2}$  and
$\text{Var}[{u_0}] = \frac{4}{\sigma_n^2}$.
Since the variance is twice the mean,  the channel message has a symmetric Gaussian distribution~\cite{richardsonBOOK2008modern}.

\subsection{Variable-to-check messages}
Consider the variable node update in (\ref{eq.VN update}). In the first  iteration of the BP decoding,
each variable node receives only a non-zero message from the channel. Hence the first set of messages passed from the variable  nodes to the check nodes follow a symmetric Gaussian PDF. The following theorem describes the  variable-to-check message exchanges in the $\ell$th, $\ell>1$,   iteration of the BP decoder.
\begin{theorem}\label{variable node update}
	The PDF of the variable-to-check message at the $\ell$th decoding iteration ($v^{(\ell)}$), is a Gaussian distribution if all check-to-variable messages ($u_i^{(\ell)}$) are Gaussian. If $u_i^{(\ell)}$s are not Gaussian then the PDF of  $v^{(\ell)}$ converges to a Gaussian distribution as the variable  node degree  tends to infinity.
\end{theorem}

\begin{IEEEproof}
The update rule at a variable node in (\ref{eq.VN update}) is the summation of the channel message and  incoming messages from check nodes ($u_i^{(\ell)}$).
Since the channel is BI-AWGN, $u_0$  follows  a symmetric Gaussian distribution. If all $u_i^{(\ell)}$s (which are mutually independent) are Gaussian, then $v^{(\ell)}$ is also Gaussian, because it is the sum of independent Gaussian random variables\cite{simonBOOK2007probability}. If $u_i^{(\ell)}$s are not Gaussian then the PDF of  $v^{(\ell)}$ converges to a Gaussian distribution as variable  node degree  tends to infinity, which directly follows from the central limit theorem~\cite{araujoBOOK1980central}.
\end{IEEEproof}
			
\begin{remark}
If $u_0$ is non-zero and has a reasonably large mean compared to $u_i^{(\ell)}$  then $u_0$ minimizes the effect of non-Gaussian PDFs coming from check nodes and  tends to sway the variable-to-check message ($v^{(\ell)}$) to be more Gaussian.  Moreover,  $v^{(\ell)}$ can be well approximated by a  Gaussian  distribution if the variable node degree is large enough.
\end{remark}

\subsection{Check-to-variable messages}
Before analyzing  the check-to-variable messages, let us first state a few useful lemmas and definitions, upon which our analysis is based.

\begin{definition}\label{lognormal}
	If the random variable $X$ is Gaussian distributed and $X = \ln (Y)$, then random variable $Y$ is said to be  lognormally distributed.
\end{definition}

\begin{lemma}[\hskip-0.5pt\cite{simonBOOK2007probability}]\label{sum of gaussain}
	If $x_1,x_2,\dots,x_n$ are independent Gaussian random variables with means $m_1,m_2,\dots,m_n$ and variances $\sigma_1^2,\sigma_2^2,\dots,\sigma_n^2$, and $\{a_i\}$ is a set of arbitrary non-zero constants, then the linear combination, $Z = \sum_{i=1}^n a_i~x_i$ follows a Gaussian  distribution with mean $\sum_{i=1}^n a_i~m_i$ and variance $\sum_{i=1}^n a_i^2~\sigma_i^2$.
\end{lemma}

\begin{lemma}\label{lognormal sum_scaled}
	Let $Y$ be a lognormal random variable. Then $(Y)^{a}$ follows a lognormal distribution, where $a \in \mathbb{Z}$.
\end{lemma}

\begin{IEEEproof}
Since $Y$ is a lognormal random variable then from Definition~\ref{lognormal}, $Y=\mathrm{e}^X$ where $X$ is Gaussian random variable. According to Lemma~\ref{sum of gaussain}, $aX$ also follows a Gaussian distribution. Thus from Definition~\ref{lognormal}, $\mathrm{e}^{aX} = (Y)^{a}$ follows a lognormal distribution.
\end{IEEEproof}

\begin{remark}\label{remark_lognormal sum} %[\hskip-0.5pt\cite{beaulieuVT2004optimal,MehtaTWC2007approximation}]
The assumption for the Gaussian approximation is that the sum of $N$ independent lognormal random variables can be well approximated by another lognormal random variable. This has been shown to be true for $N=2$~\cite{beaulieuVT2004optimal}.
\end{remark}

Now consider the check node update in (\ref{eq.CN update}) at the $\ell$th decoding iteration.
\begin{remark}\label{check node update}
	The PDF of the check-to-variable message at the $\ell$th decoding iteration ($u^{(\ell)}$) is  a Gaussian distribution provided that the variable-to-check messages are approximately Gaussian and  reasonably reliable\footnotemark[3]\footnotetext[3]{Since the all zero codeword is transmitted, reasonably reliable messages suggest that majority of $v^{(\ell)}$'s take large positive values (i.e.,  $v^{(\ell)}$ has a large mean).},  and the degrees of the check nodes are small\footnotemark[4]\footnotetext[4]{Following Remark~\ref{remark_lognormal sum}, by ``small check node degree'' we mean the check node degree equals two.}.
\end{remark}

%\begin{IEEEproof}
Consider the check node with degree $d_c$.  We can rewrite (\ref{eq.CN update}) as follows:
\begin{align}
u_e^{(\ell)}
&=\left(\prod_{j=1}^{d_c-1}\text{sign}(v_j^{(\ell)})\right)~\underbrace{~\varphi\left( \overbrace{~\sum_{j=1}^{d_c-1}~ \underbrace{~\varphi(~v_j^{(\ell)}~)}_\text{Step 1}~}^\text{Step 2}\right)~}_\text{Step 3}
\label{check update 2}
\end{align}
where we define	$\varphi(x) =\log \left(\tanh\frac{x}{2}\right) =\log\left(\frac{ \mathrm{e}^{x} -1}{ \mathrm{e}^{x}+1}\right)  \text{for} ~x \geqslant 0 $ and note that 
\begin{align*}
\varphi^{-1}(x) &=\log\left(\frac{ \mathrm{e}^{x} +1}{ \mathrm{e}^{x}-1}\right) =\varphi(x).
\end{align*}
Suppose  $v_j^{(\ell)}$s are reasonably reliable. Using Taylor series expansion,  $\varphi(v)$ can be expressed as 
\begin{align*}
\varphi(v) &=\log\left(\frac{ \mathrm{e}^{v} -1}{ \mathrm{e}^{v}+1}\right) = 2(\mathrm{e}^{v})^{-1} + \frac{2}{3}(\mathrm{e}^{v})^{-3} + \frac{2}{5}(\mathrm{e}^{v})^{-5} + \dots
\end{align*}
For  simplicity, we omitted the indices  $j$, $\ell$.  	
Since $v$ follows an approximate Gaussian distribution, according to Definition~\ref{lognormal}, $\mathrm{e}^v$ follows an approximate lognormal distribution and  from  Lemma~\ref{lognormal sum_scaled} $(\mathrm{e}^v)^{b_i}$, $b_i\in \mathbb{Z}$ follows an approximate lognormal distribution. Using the assumption that, the  sum of a set of independent lognormal random variables is approximately lognormal when the set size is small (see Remark~\ref{remark_lognormal sum}), $\varphi\big( v \big)$  (in step 1) follows an approximate lognormal distribution.  This is because, when $v$ is large with high probability, the higher-order terms in the Taylor series expansion of $\varphi(v)$ are insignificant compared to the first few terms. Next, the $\varphi\big( v_j^{(\ell)}\big)$s are mutually independent. Thus, according to  Remark~\ref{remark_lognormal sum}, $\sum_{j=1}^{d_c-1}\varphi\big( v_j^{(\ell)}\big)$ (in step 2) follows an approximate lognormal distribution when  the check node degree is small. Finally, from Definition~\ref{lognormal}, $\varphi \left(\sum_{j=1}^{d_c-1}\varphi\big( v_j^{(\ell)}\big) \right)$ (in step 3) will follow an approximate Gaussian distribution if the result of step 2 is lognormally distributed.	
%\end{IEEEproof}	
\begin{remark}\label{remark_lognormal sum_not_correct} 
	The assumption that, the sum of $N$ independent lognormal random variables can be well approximated by another lognormal random variable, is not true when $N$ is large. 
\end{remark}

Using the above results, we  investigate  the accuracy of  Gaussian approximations to full-DE of  low rate MET-LDPC codes, with  punctured and degree-one variable nodes. These are the cases where MET-LDPC  codes are most beneficial. We also evaluate full-DE simulations for these codes to measure how close the  actual message PDF is (under the  full-DE) to a Gaussian PDF using the Kullback-Leibler (KL) divergence~\cite{coverBOOK2012elements} as our measure.  A small value of  KL divergence  indicates that actual PDF is close to a Gaussian PDF. We calculate the KL divergence between 1) the actual message PDF (under the  full-DE) and a  Gaussian PDF with the same mean and variance to check whether it follows a  Gaussian distribution, 2) the actual message PDF (under the  full-DE) and a symmetric Gaussian PDF with the same mean to check whether it follows a symmetric Gaussian distribution. \\

\subsubsection{Low SNR} \label{Low-rate codes}
In the case of standard LDPC codes, it has been observed that~\cite{fuICC2006gaussian,ardakaniCOM2004more,xieISIT2006accuracy} the check-to-variable messages  significantly deviate from a symmetric Gaussian distribution at low signal-to-noise ratios (SNR), even if the variable-to-check messages are close to a Gaussian distribution. Thus Gaussian approximations based on single-parameter  models do not  perform well for the codes at  low SNRs. Here we explain the reason behind  this, based on the assumptions required for Gaussian approximations to be accurate.

%figure for low-rate problem low SNR
\begin{figure}[t]
	\centering
	\includegraphics[width=1\linewidth]{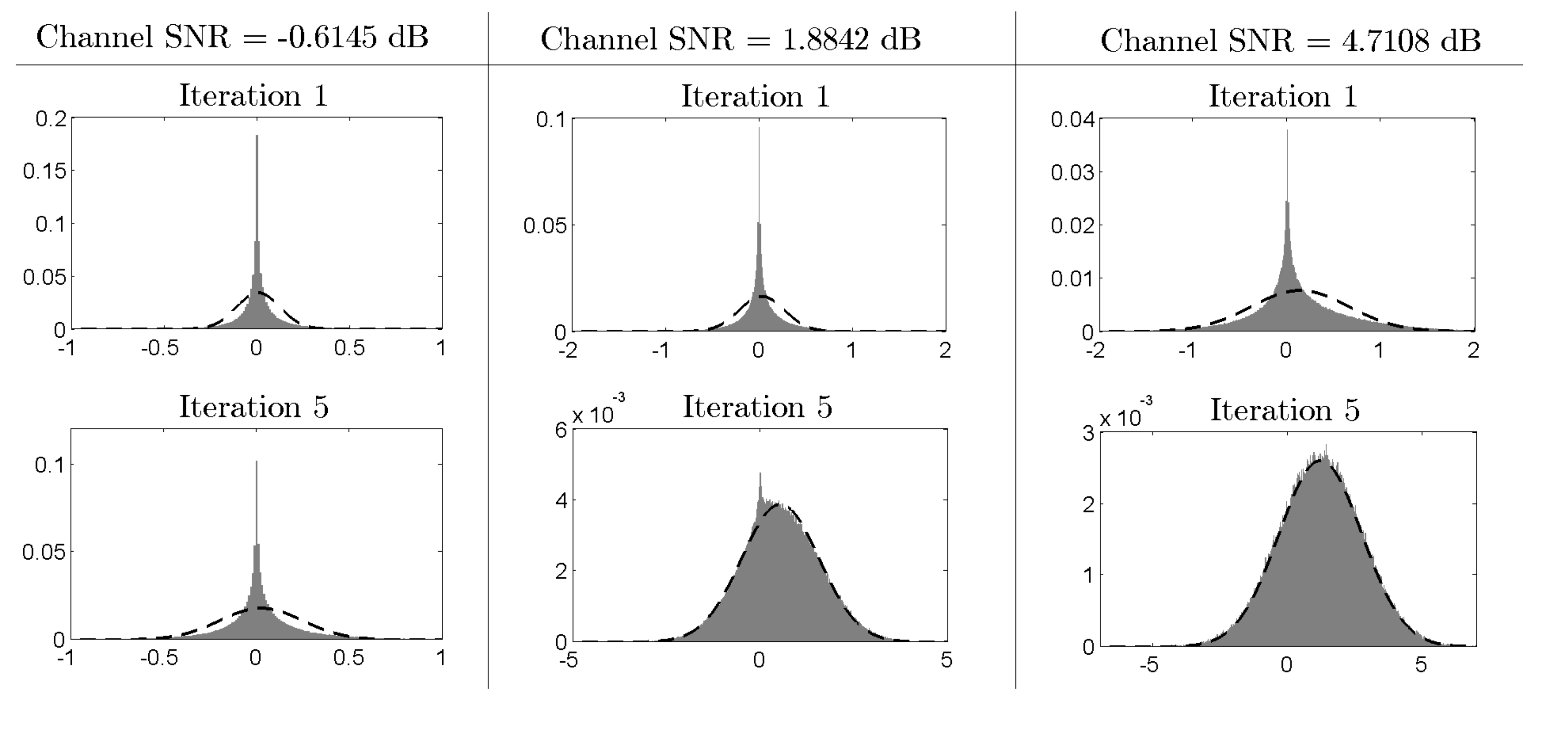}
	\vspace*{-1em}
	\caption{Output PDF of  check-to-variable  messages from edge-type two  of  rate $1/10$  MET-LDPC code with $L(\boldsymbol{r},\boldsymbol{x}) = 0.1 r_1 x_1^3 x_2^{20} + 0.025 r_1 x_1^{3} x_2^{25}+ 0.875 r_1 x_3,~ R(\boldsymbol{x}) = 0.025 x_1^{15} + 0.875 x_2^3x_3$ is compared with the symmetric Gaussian PDF of the same mean.}
	\label{fig:pdf_R}
	%\vspace*{-2em}
\end{figure}

%figure for low-rate problem (KL divergence)
\begin{figure}[t]
	\centering
	\includegraphics[width=1\linewidth]{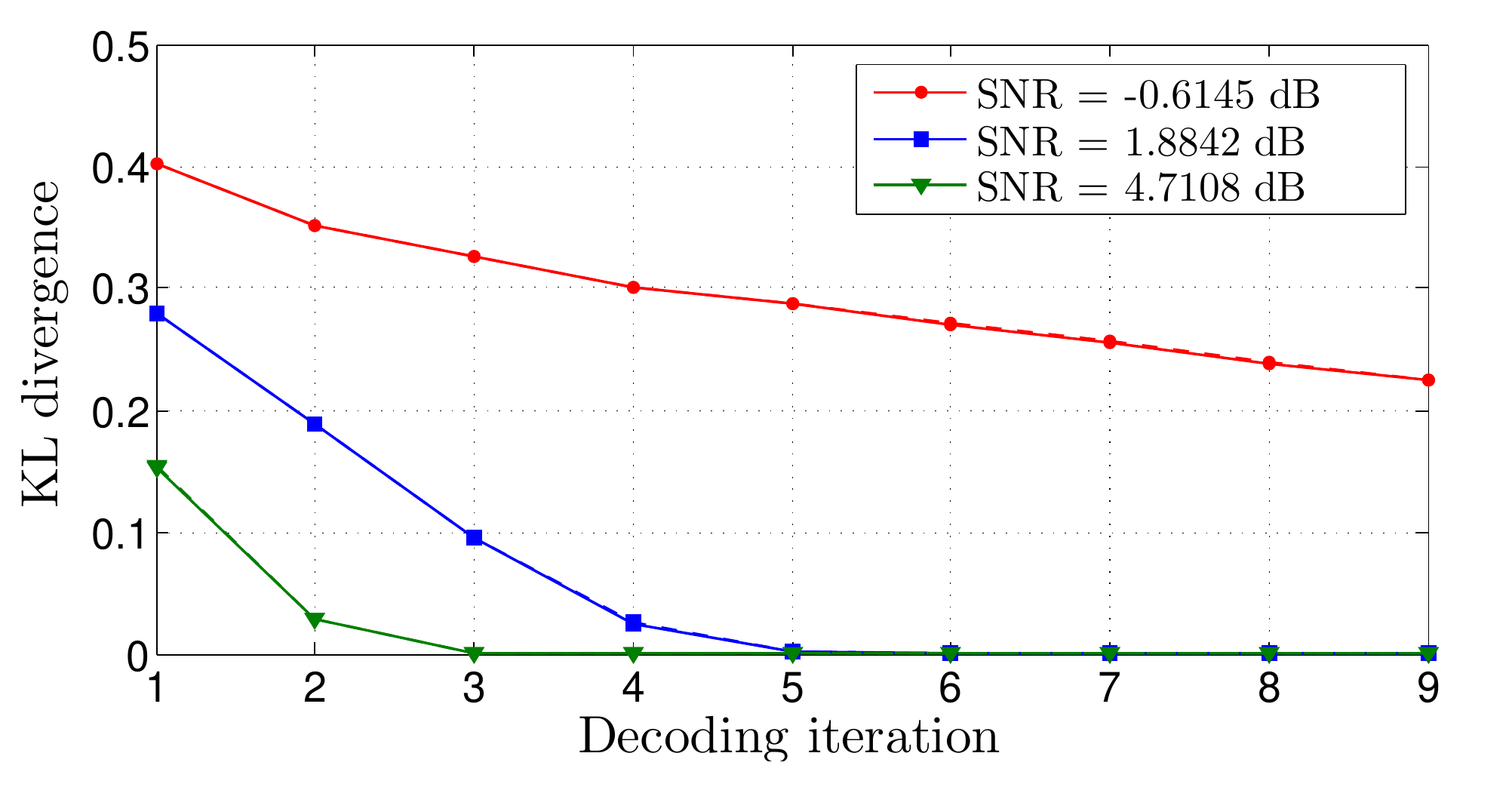}
	\vspace*{-1em}
	\caption{KL divergence of check-to-variable message  PDF from edge-type two to the corresponding Gaussian PDF (solid line) and to the corresponding symmetric Gaussian PDF (dotted line) of rate $1/10$ MET-LDPC code with $L(\boldsymbol{r},\boldsymbol{x}) = 0.1 r_1 x_1^3 x_2^{20} + 0.025 r_1 x_1^{3} x_2^{25}+ 0.875 r_1 x_3,~ R(\boldsymbol{x}) = 0.025 x_1^{15} + 0.875 x_2^3x_3$. }
	\label{fig:KL_low rate}
	%\vspace*{-2.5em}
\end{figure}

According to Remark~\ref{check node update}, the PDF of the check-to-variable messages ($u^{(\ell)}$) can be well approximated by a Gaussian PDF, if the variable-to-check messages ($v_j^{(\ell)}$s) are reasonably reliable given that $v_j^{(\ell)}$s are approximately Gaussian and check node degrees are small. At low SNR, the initial  $v_j^{(\ell)}$s are not reasonably reliable. Thus $u^{(\ell)}$ may not follow a Gaussian distribution in early decoding iterations.  However, if the SNR is above the code threshold,  the decoder  converges to zero error probability as decoding iterations proceed, thus the PDF of $v_j^{(\ell)}$ moves to right  and  the $v_j^{(\ell)}$s become more reliable. Hence $u^{(\ell)}$ may  follow a Gaussian distribution at later decoding iterations.

To illustrate this via an example, we plot the actual message PDFs in the BP decoding and the KL divergence between the PDF of check-to-variable message from edge-type two and the corresponding symmetric Gaussian PDF for a rate $1/10$ MET-LDPC code in Figs.~\ref{fig:pdf_R} and~\ref{fig:KL_low rate}, respectively. It is clear from Fig.~\ref{fig:KL_low rate} that the KL divergence  at a low SNR has a larger value than that at high SNRs. This shows that $u^{(\ell)}$  significantly deviates from a  Gaussian distribution  when the SNR is low. We  can also see from  Figs.~\ref{fig:pdf_R} and~\ref{fig:KL_low rate} that $u^{(\ell)}$ follows a Gaussian distribution at later decoding iterations, and  the lower the SNR the more decoding iterations are required  for this to happen.  Based on these observations  we  claim that single-parameter Gaussian approximations may not be a good approximation to DE at low SNRs.\\

\subsubsection{Large check node degree}
%figure for low-rate problem and high check node degree
\begin{figure}[t]
	\centering
	\includegraphics[width=1\linewidth]{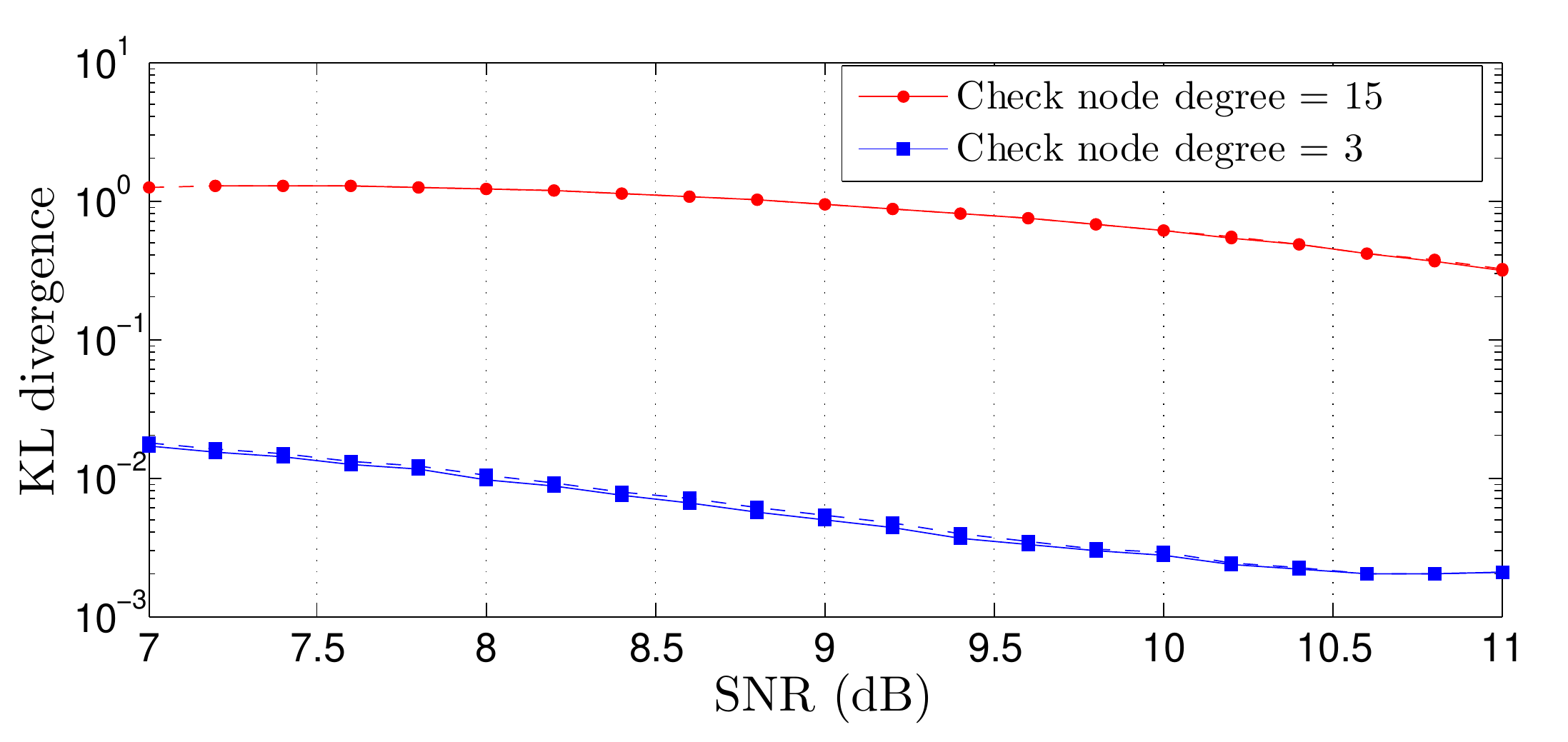}
	\vspace*{-1em}
	\caption{KL divergence of check-to-variable message PDF to the corresponding Gaussian PDF (solid line) and to the corresponding symmetric Gaussian PDF (dotted line) of rate $1/10$ MET-LDPC code with $L(\boldsymbol{r},\boldsymbol{x}) = 0.375 r_1 x_1^2 + 0.625 r_1 x_1^{6}, ~R(\boldsymbol{x}) = 0.15 x_1^{15} + 0.75 x_1^3$ at the first decoding iteration.}
	\label{fig:SNRandcheckdegree}
	%\vspace*{-2.5em}
\end{figure}

It has been observed~\cite{fuICC2006gaussian,xieISIT2006accuracy,ardakaniCOM2004more} that  the check-to-variable messages   significantly deviate from a symmetric Gaussian distribution when the check node degree is large, even if the variable-to-check messages are close to a Gaussian distribution. Thus single-parameter  Gaussian approximation  models do not  perform well for the standard LDPC codes  with large  check node degrees. Here we explain the reason behind this, based on the assumptions required for Gaussian approximations to be accurate.

According to Remark~\ref{check node update}, the PDF of the check-to-variable messages ($u^{(\ell)}$s) can be well approximated by a Gaussian PDF, when  check node degrees are small given that $v_j^{(\ell)}$s are approximately Gaussian and reasonably reliable.  The assumption in step 2 (see Remark~\ref{remark_lognormal sum}), that is the sum of $N$ independent lognormal random variables can be well approximated by another lognormal random variable, is clearly not true if the check node degree is large (see Remark~\ref{remark_lognormal sum_not_correct}). Thus $u^{(\ell)}$ may not follow a Gaussian distribution for larger the check node degrees.

To evaluate the combined effect of the SNR and the check node degree, we plot the  KL divergence of check-to-variable message PDFs to the corresponding symmetric  Gaussian PDFs  for a rate $1/10$ MET-LDPC code at the first decoding iteration for different SNRs and different check node degrees  in Fig.~\ref{fig:SNRandcheckdegree}.  Our simulations show that with a large check node degree of 15, the KL divergence is large. Based on this we  claim that single-parameter Gaussian approximations may not be a good approximation to DE for codes with large check node degrees. \\

\subsubsection{The effect of punctured variable nodes} \label{punctured variable nodes}
One of the modifications of MET-LDPC codes over standard LDPC codes is the addition of punctured variable nodes to improve the code threshold (a different use of puncturing  than its typical use to increase the rate).  We observe that these punctured nodes have a significant impact on the accuracy  of the Gaussian approximation of both variable-to-check messages and check-to-variable messages.  According to Theorem~\ref{variable node update}, if $u_i^{(\ell)}$s are not Gaussian, the PDF of the variable-to-check messages ($v^{(\ell)}$) converges to a Gaussian distribution as the variable node degree tends to infinity. In the case of  punctured nodes, $u_0$ equals zeros as punctured bits are not transmitted through a channel. Hence in  punctured variable nodes, $v^{(\ell)}$ is equivalent to the sum of $u_i^{(\ell)}$s only, which are  heavily non Gaussian at early decoding iterations. Thus if the variable node degree is not large enough, then  $v^{(\ell)}$ from punctured variable nodes may  not follow a Gaussian distribution at  early decoding iterations.

%figure for punctured nodes (KL divergence)
\begin{figure}[t]
	\centerline{
		\subfigure[Variable-to-check message]{\includegraphics[width=1\linewidth]{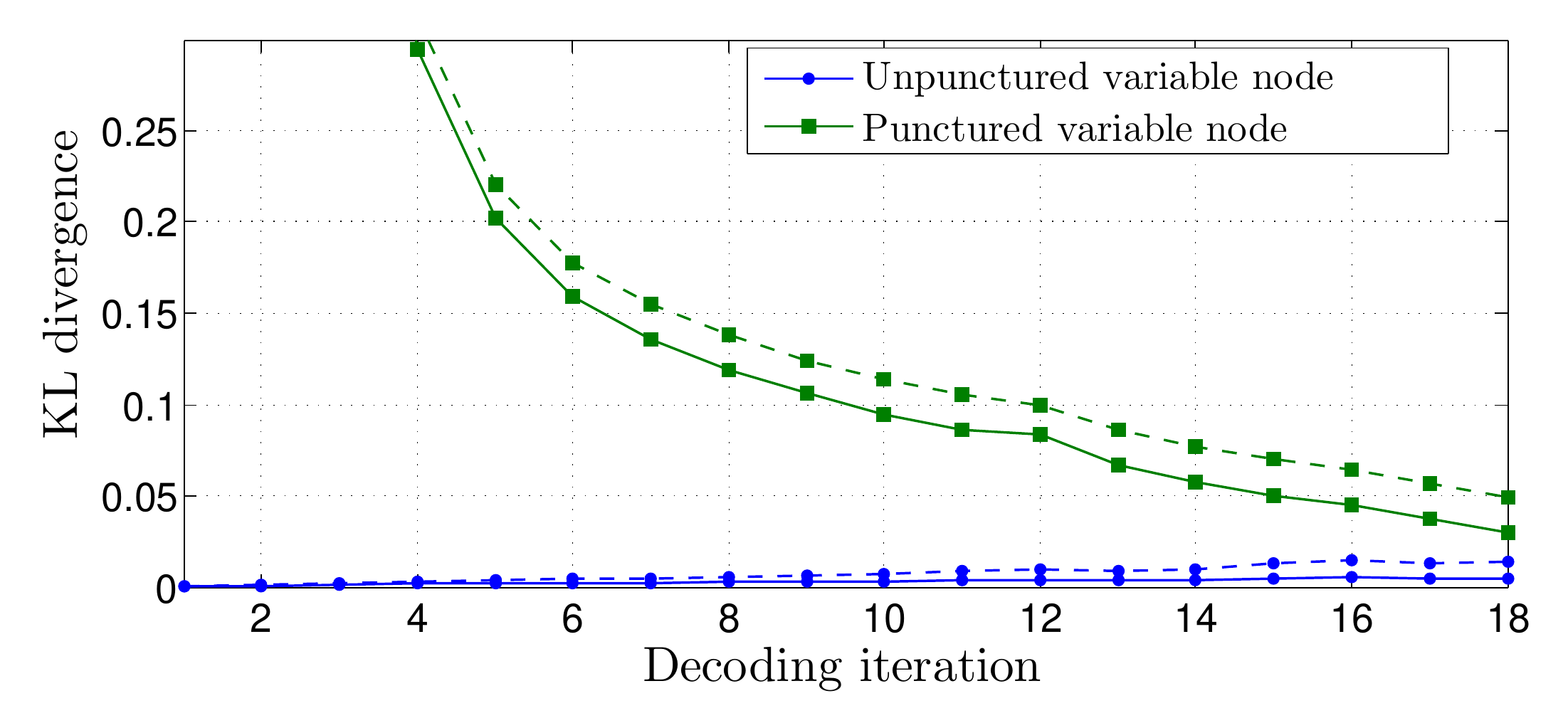} \label{fig:KL_varibale_p}}}
		\vfill
	\centerline{
		\subfigure[ Check-to-variable message]{\includegraphics[width=0.93\linewidth]{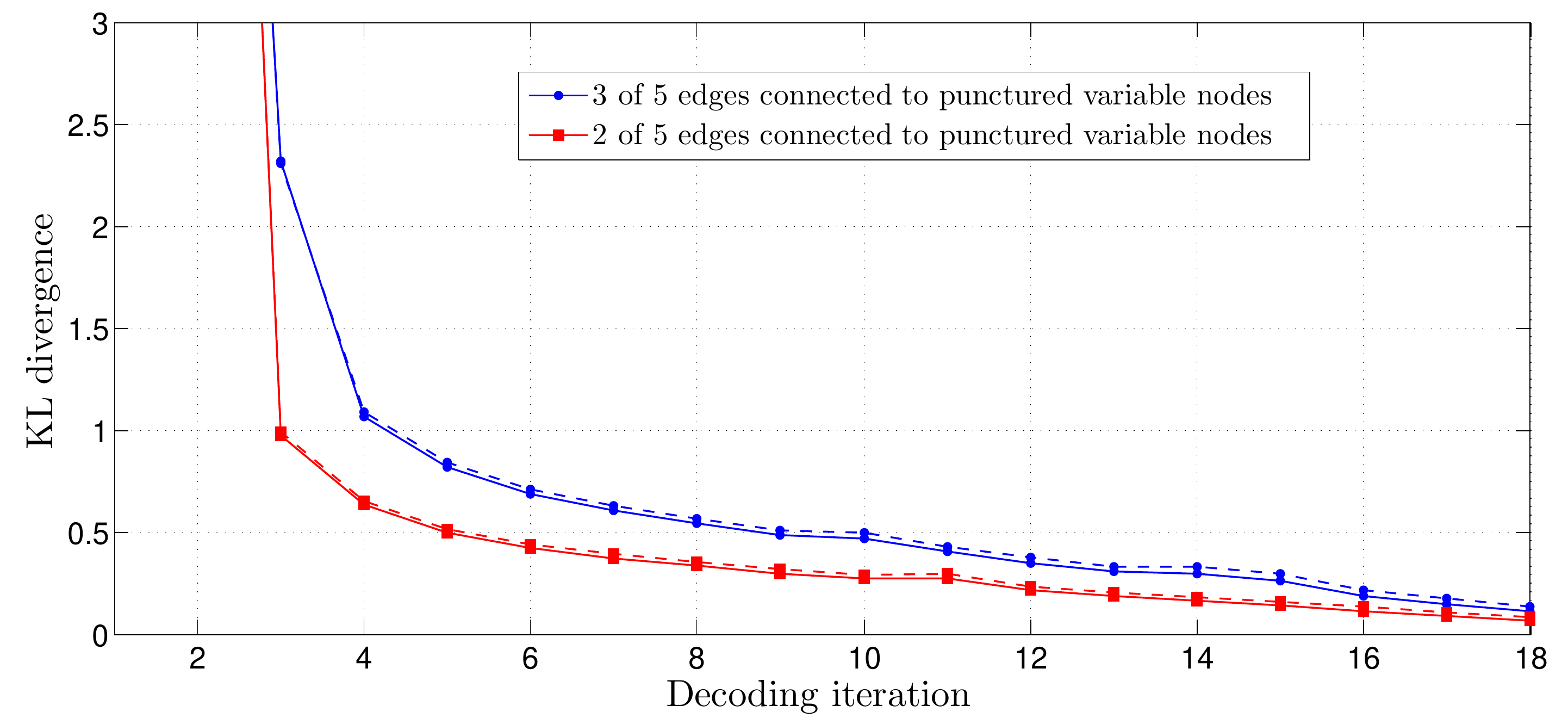}\label{fig:KL_check_p}}}
%	\vspace{-1em}
	\caption{KL divergence of variable-to-check and check-to-variable message PDFs to the corresponding Gaussian PDFs (solid line) and to the corresponding symmetric Gaussian PDFs (dotted line) of rate $1/2$ MET-LDPC code with $L(\boldsymbol{r},\boldsymbol{x}) = 0.5 r_1 x_1^3 x_2^3 + 0.5 r_1 x_1^{3} + 0.5 r_0x_2^3,~ R(\boldsymbol{x}) = x_1^3x_2^3$,  fed with channel noise standard deviation 0.05 below the code threshold.}
	\label{fig:KL_P} 
	%\vspace{-2em}
\end{figure}

The punctured variable nodes adversely affect the check-to-variable messages as well. According to Remark~\ref{check node update}, the PDF of the check-to-variable messages ($u^{(\ell)}$) can be well approximated by a Gaussian PDF, if the variable-to-check messages ($v_j^{(\ell)}$s) are well approximated by a Gaussian PDF. This is because,
$\varphi(v)$ in step 1 of (\ref{check update 2})  follows an approximate lognormal distribution only if $v^{(\ell)}$ is following a Gaussian distribution.  Since  $v^{(\ell)}$ from a punctured variable node does not follow a Gaussian distribution,  $u^{(\ell)}$ also may not follow a  Gaussian distribution.  The end result is that punctured nodes reduce the validity of the Gaussian approximation for $v^{(\ell)}$  and $u^{(\ell)}$.

To illustrate the effect of punctured nodes via an example, we plot the KL divergence of variable-to-check and check-to-variable messages to the corresponding symmetric Gaussian PDFs of rate a $1/2$ MET-LDPC code with punctured nodes in Fig.~\ref{fig:KL_P}.  It is clear from Fig.~\ref{fig:KL_varibale_p} that the KL divergence  of the variable-to-check message to the corresponding symmetric Gaussian PDF from  punctured nodes has a larger value than that from the unpunctured nodes in the same code. Fig.~\ref{fig:KL_check_p} shows the corresponding effect on the KL divergence of the check-to-variable messages.  Furthermore,  the decrease of the KL divergence  with  decoding iterations in Fig.~\ref{fig:KL_P}  implies that  $v^{(\ell)}$ and  $u^{(\ell)}$ are following a Gaussian distribution at later decoding iterations. However in general,  to become Gaussian it takes more decoding iterations than typical for a code without punctured nodes.

\subsubsection{The effect of degree-one variable nodes} \label{degree-one variable nodes}
One of the advantages of the MET-LDPC codes is  the addition of degree-one variable nodes to improve the code threshold. However we observe that   degree-one variable nodes can affect the Gaussian approximation for the check-to-variable messages. Here we explain the reason behind this, based on the assumptions required for Gaussian approximations to be accurate.

According to Remark~\ref{check node update}, the PDF of the check-to-variable messages ($u^{(\ell)}$) can be well approximated by a Gaussian PDF, if the variable-to-check messages ($v_j^{(\ell)}$s) are reasonably reliable given that $v_j^{(\ell)}$ are Gaussian. Even though  $v_j^{(\ell)}$s received from edges  connected to degree-one variable nodes 
are Gaussian, they  are not reasonably reliable. This is because variable nodes of degree-one never update their $v_j^{(\ell)}$s as they do not receive information from more than one neighboring check node. So the $u^{(\ell)}$s may not follow Gaussian distribution.  Consequently, this may reduce the validity of the Gaussian approximation to DE for MET-LDPC codes.

Similarly any check node that receives input messages from a degree-one variable node never outputs a  distribution  with an infinitely large mean in any of its edge messages updated with information from  degree-one variable nodes. This is because, if $x_1 \dots x_n$ are a set of independent random variables, then
${\mathcal{L}}(x_1)\oplus{\mathcal{L}}(x_2) \oplus \dots \oplus  {\mathcal{L}}(x_n) = {\mathcal{L}}(x_1 \oplus x_2 \oplus \dots \oplus x_n),$
and $\mathcal{L}(x_1 \oplus x_2 \oplus \dots \oplus x_n) \stackrel{p}{\rightarrow} \mathcal{L}(x_1)$ as $\min_{2 \leq i \leq n} \text{E}[\mathcal{L}(x_i)] \rightarrow \infty$~{\protect\cite[pages 735-738]{moonBOOK2005Error}}, as is the case with degree-one variable nodes.  We observed through the simulations that the degree-one variable nodes  have a small impact on the accuracy  of the Gaussian approximation of both variable-to-check messages and check-to-variable messages.

%***********************************************************************************************
%***************Section V - Hybrid density evolution********************************************
%***********************************************************************************************
\vspace*{1em}
\section{Hybrid density evolution for MET-LDPC codes} \label{Hybrid DE}
All the Gaussian approximations we discussed in Section~\ref{Gaussian app} are  based on the assumption that the PDFs of the variable-to-check and the check-to-variable messages can be well approximated by  symmetric Gaussian distributions. This assumption is quite accurate at the later decoding iterations, but least accurate in the early decoding iterations particularly at low SNRs or with punctured variable nodes or with large check node degrees as we have observed in Section~\ref{validity of GA}. Making the assumption of symmetric Gaussian distributions at the beginning of the DE calculation produces large errors between the estimated and  true distributions. Even when the true distributions do become Gaussian, the approximations give  incorrect Gaussian distributions due to the earlier errors. These errors propagate throughout the DE calculation and cause significant errors in the final code threshold result for MET-LDPC codes as we will see in Section~\ref{code design GA}. Through  simulations, we observed that, when the channel SNR is above the code threshold, the PDFs of the node messages (i.e., variable-to-check and the check-to-variable  messages)  eventually do become symmetric Gaussian distributions as decoding iterations proceeds. This implies that  making the assumption of symmetric Gaussian distributions in the  later decoding iterations of the DE calculation is reasonable. This motivates us to propose a hybrid density evolution (hybrid-DE) algorithm for MET-LDPC codes  which is a combination of the  full-DE and  the mean-based Gaussian approximation (Approximation 1). The key idea in hybrid-DE is that  we do not assume  that the node messages are symmetric Gaussian at the beginning of the DE calculation, i.e., hybrid-DE method initializes the node message PDFs using the  full-DE and then switches to  the Gaussian approximation.

There are two options for switching  from the full-DE to the Gaussian approximation.  As  option one, we can impose a limitation for the number  of maximum full-DE  iterations in hybrid-DE, in which we do few full-DE  iterations and then switch to the Gaussian approximation. Although this is the simplest option, it gives a nice trade-off between accuracy  and  efficiency of threshold computation as shown in Fig.~\ref{fig:MET_threshold_it}.   The second option is that  we can switch  from full-DE to the Gaussian approximation  after that the PDFs for the node  messages   become nearly symmetric Gaussian.  The KL divergence~\cite{coverBOOK2012elements} can be  used to check whether a  message PDF is close to a symmetric Gaussian distribution.  Thus, as the second option, we can impose a limitation for the KL divergence between the actual node message PDF and a symmetric Gaussian PDF. This is a more accurate way of switching than option one. Because the value of the KL divergence  depends on the shape of the PDF of the node messages, thus the switching point is changing appropriately with the condition (such as SNR, code rate) we are looking at.

Each option has its own pros and cons. For instance,  if the channel SNR is well above the code threshold, the node messages can be close to a symmetric Gaussian distribution before the imposed limit in option one for the full-DE iterations. Thus  we are doing extra full-DE  iterations that are not necessary. This  reduces the benefit of hybrid-DE by adding extra run-time. In such a situation, we can introduce the second option (i.e., KL divergence limit) in addition to reduce run-time by halting the full-DE iterations once the PDF is sufficiently Gaussian. On the other hand,   if the channel SNR is below the code threshold, the decoder never converge to a zero-error probability as decoding iterations proceed, thus  node messages may not ever   follow a symmetric Gaussian distribution. This makes the option two hybrid-DE always remain at full-DE,  as it never meets the target KL divergence limit.  Thus  forcing a limitation  for the number of full-DE  iterations (i.e., option one) is  required in order to improve the run-time of hybrid-DE.
Because of these reasons, we can introduce both options to the hybrid-DE where option one acts as a hard limit and option two acts as a soft limit.  This is a particularly beneficial way to do  the trade-off between accuracy  and  efficiency when computing the code threshold.  We  found that it is possible to impose both options in the hybrid-DE to significantly improve computational time without significantly reducing  the accuracy of the threshold calculation.

%table for complexcity of approximations

\begin{table*}[t]
	\renewcommand{\arraystretch}{1.5}
	\caption{Floating point (FP) operations per edge, based on an average edge degree of variable node (VN), $\bar{d_v}$ and an average edge degree of check node (CN), $\bar{d_c}$}
	\label{Table : FP values}
	\centering	
	%\begin{tabular}{|C{1.6cm}|C{0.5cm}|C{0.8cm}|C{0.5cm}|C{0.8cm}|C{0.5cm}|C{1cm}|C{0.8cm}|C{0.8cm}|C{1.5cm}|C{2cm}|}
	\scriptsize
	\begin{tabular}{|l|c|c|c|c|c|c|c|c|c|c|}
		\hline
		\multicolumn{1}{|c|}{FP}  &\multicolumn{2}{c|}{\multirow{2}[2]{*}{MET-DE}} &  \multicolumn{2}{c|}{App. 1} &  \multicolumn{2}{c|}{App. 2} &  \multicolumn{2}{c|}{App. 3} &  \multicolumn{2}{c|}{\multirow{2}[2]{*}{Hybrid-DE\footnotemark[4]}}   \\
		\multicolumn{1}{|c|}{\multirow{2}[1]{*}{ operation}} & \multicolumn{2}{c|}{} & \multicolumn{2}{c|}{(Mean)}& \multicolumn{2}{c|}{(BER)}&  \multicolumn{2}{c|}{(RCA)}  & \multicolumn{2}{c|}{} \\
		\cline{2-11}
		\multicolumn{1}{|c|}{ }& VN & CN & VN & CN & VN & CN & VN & CN & VN & CN \\
		\hline
		\hline
		%&  &  & & & & & & &&\\
		Sums &  &  & $\bar{d_v}$ & $\bar{d_c}$ & $\bar{d_v}$ & $2\bar{d_c}+1$ & $\bar{d_v}$ & $\bar{d_c}-1$ & $(1-\alpha)\bar{d_v}$ & $(1-\alpha)\bar{d_c}$ \\
		%\hline
		Multiplications &  &  &  &  & 2 &  &  &  &  &  \\
		%\hline
		Lookup-tables &  &  &  & $\bar{d_c}$ &  &  &  $\bar{d_v}-1$ & $\bar{d_c}-1$  &  & $(1-\alpha)\bar{d_c}$ \\
		%\hline
		Exponentials &  &  &  & $\bar{d_c}-1$ & 1 & $\bar{d_c}-1$ & & &  & $(1-\alpha)(\bar{d_c}-1)$ \\
		%\hline
		Q-functions &  &  &  &  & 1 &  &  &  &  &  \\
		%\hline
		Convolutions & $\bar{d_v}$ & $\bar{d_c}-1$ &  &  &  &  &  &  & $\alpha\bar{d_v}$ & $\alpha(\bar{d_c}-1)$ \\
		\hline
	\end{tabular}
	
	\begin{tablenotes}
		\small
		\item {$^4\alpha$ is the percentage of  MET-DE  iterations.}
	\end{tablenotes}
	%\vspace*{-3em}
\end{table*}
\normalsize

Throughout this paper, the check-to-variable message with the largest check node degree is chosen to check the KL divergence because the most significant errors relating to the estimation of the PDF  occurs at  large degree check nodes  as we observed in Section~\ref{validity of GA}.  While running,  the DE algorithm periodically  calculates the KL divergence between the actual message PDF (under the  full-DE) and a symmetric Gaussian PDF with the same mean for the selected check node message.
The hybrid-DE continues using the  full-DE until the KL divergence is smaller than a predefined target KL divergence or  a predetermined maximum number of  full-DE  iterations is reached when it then switches to a Gaussian approximation DE. Thus, we can trade-off accuracy for efficiency of the hybrid-DE method by varying the target KL divergence and/or  the maximum number of  full-DE  iterations.

%***********************************************************************************************
%***************Section VI - Code design********************************************
%***********************************************************************************************

\section{Implication of  Gaussian approximations for code design} \label{code design GA}

\subsection{Threshold comparison of density evolution using Gaussian approximations} \label{threshold}

Table~\ref{Table : FP values} gives the number of floating point operations per edge per decoding iteration for each of the DE algorithms.  We do not show the overhead operations (such as computing the KL divergence) that do not occur during the DE iterations in Table~\ref{Table : FP values}. We have found that these operations make only a small contribution to the overall overhead. The relative complexity and accuracy of each approach will depend on the size of the lookup table chosen (for Approximations 1 and 3) and the number of quantization points chosen to sample the PDF (for  full-DE) or for Q-function evaluation (for Approximation 2).

\begin{figure}[t]
	\centering
	\includegraphics[width=1\linewidth]{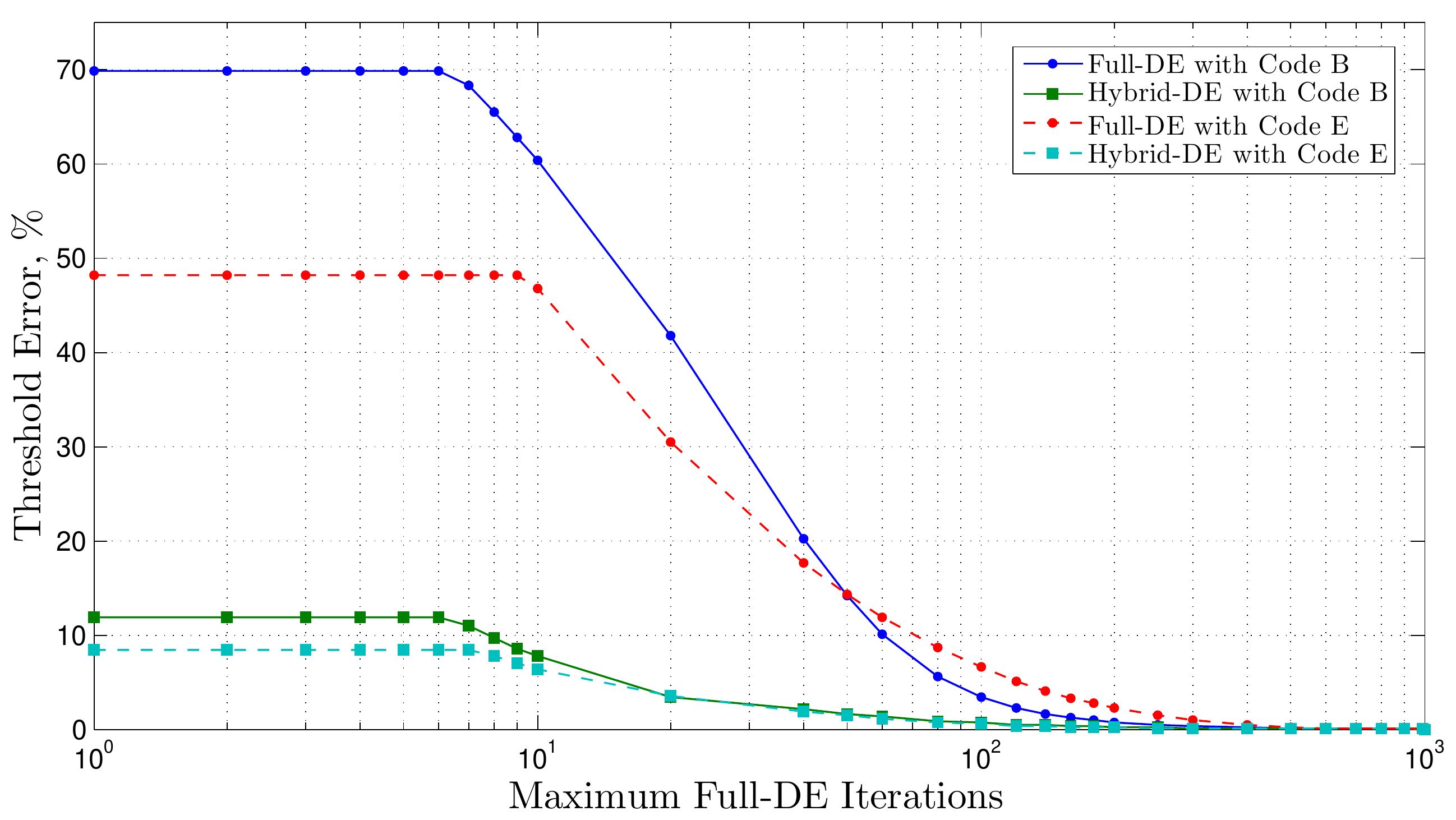}
	\vspace*{-1em}
	\caption{Percentage of threshold error$^6$, with respect to the full-DE threshold with $1000$ decoding iterations,  for different maximum number of full-DE iterations. After the maximum number of full-DE iterations is reached, full-DE stops, while hybrid-DE continues with a Gaussian approximation (Approximation 1) for up to $1000$ decoding iterations. No KL divergence limit is set for hybrid-DE.}
	\label{fig:MET_threshold_it}
	%\vspace{-2em}
\end{figure}

\begin{figure}
	\centering
	\includegraphics[width=1\linewidth]{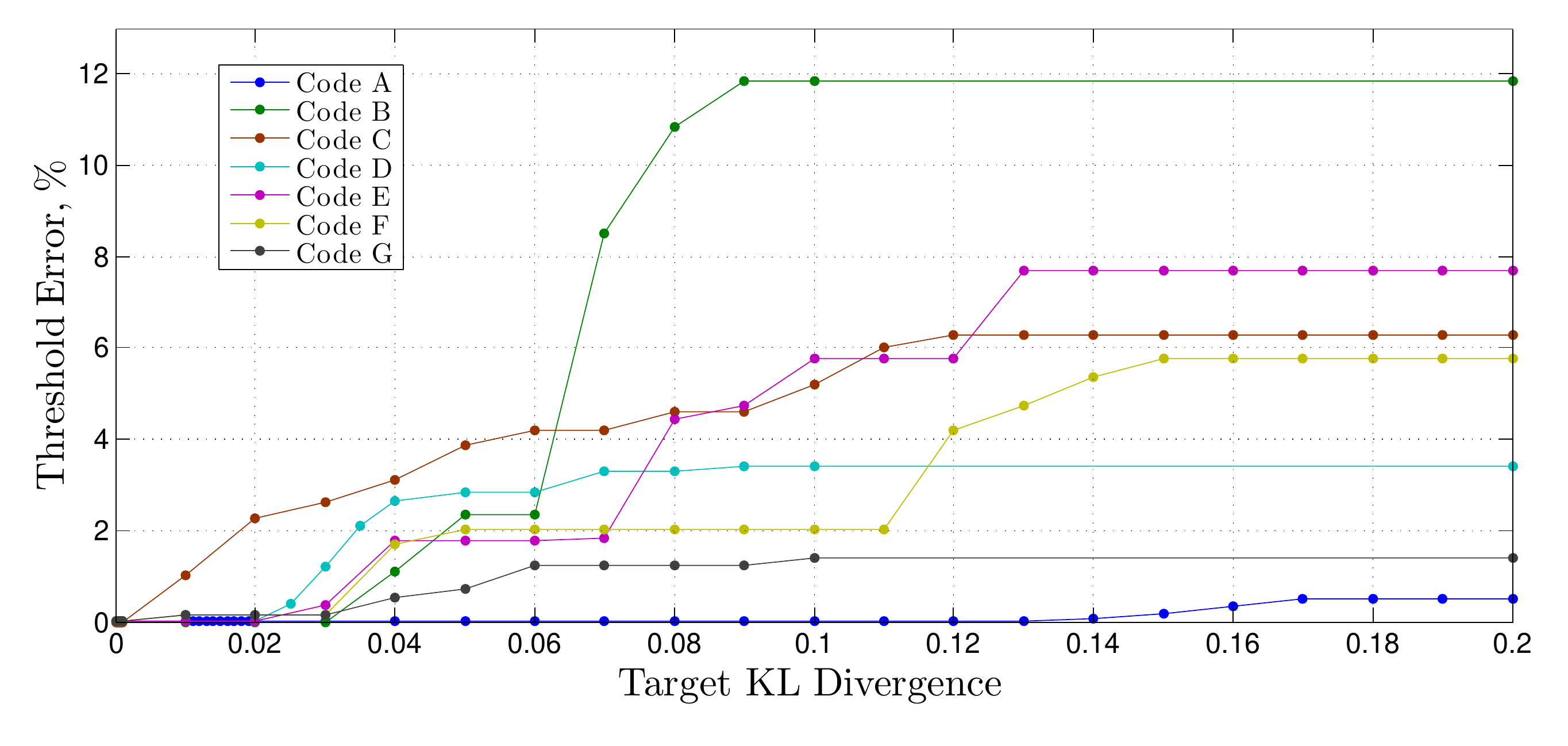}
	\vspace*{-1em}
	\caption{Percentage of threshold error$^6$, with respect to  the full-DE threshold with $1000$ decoding iterations,  for different target KL divergence limits. No maximum number of full-DE iterations is set. The hybrid-DE algorithm swaps to Gaussian iterations (Approximation 1) only once the target KL divergence is met. }
	\label{fig:MET_threshold_KL}
	%\vspace{-2em}
\end{figure}

\begin{figure}[t]
	\centering
	\includegraphics[width=1\linewidth]{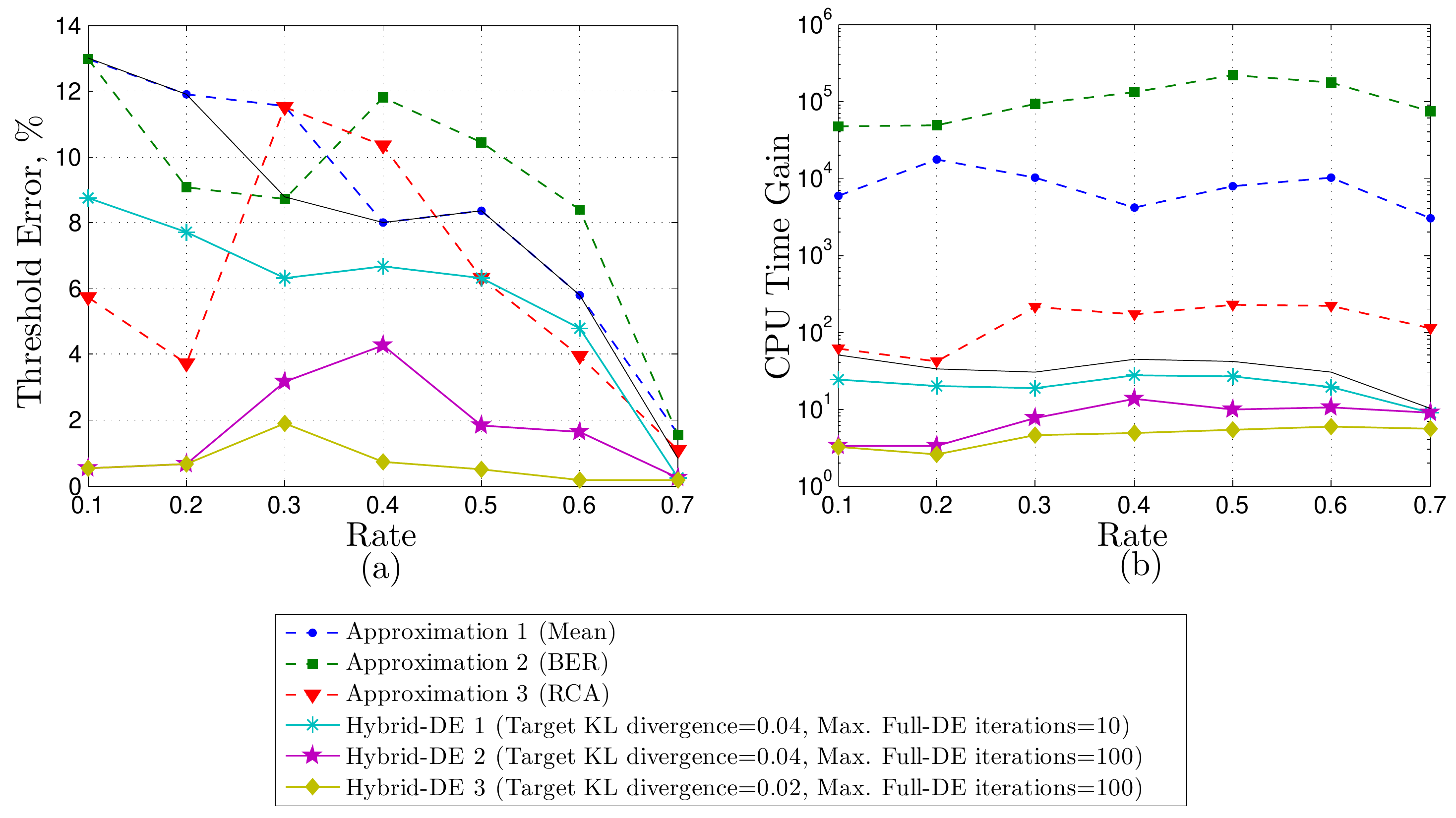}
	\vspace*{-1em}
	\caption{(a) Percentage of threshold error$^6$, with respect to the full-DE threshold with $1000$ decoding iterations. (b) CPU time gain for one DE calculation$^7$, with respect to  the  full-DE with $1000$ decoding iterations,  when channel noise standard deviation is 0.01 below the code threshold. Rates 0.1 to 0.7 correspond to  codes A to G in Table~\ref{Table : MET_examples_codes} in Appendix. }	
	\label{Fig.MET_threshold_time}
	%\vspace{-2em}
\end{figure}

In Figs.~\ref{fig:MET_threshold_it} to~\ref{Fig.MET_threshold_time} we compare the percentage of threshold error  and CPU time gain, with respect to  the threshold and CPU time obtained from full-DE with $1000$ decoding iterations, for MET-LDPC codes of different rates.  We select these MET-LDPC  code structures (code A-G in Table~\ref{Table : MET_examples_codes} in Appendix) such that they contain degree-one and punctured variable nodes in oder to emphasize the benefits of hybrid-DE over Gaussian approximations.  We specified $9800$ quantization points per PDF for  full-DE  and lookup table sizes of $10001$ and $38302$ for Approximations 1 and 3 respectively.  This is  because we found that assigning  smaller lookup tables (for Approximations 1 and 3) and smaller  number of quantization points  (for  full-DE) reduces the accuracy of the threshold calculation. 

Figs.~\ref{fig:MET_threshold_it} and~\ref{fig:MET_threshold_KL}  present  the effect of  the maximum number of full-DE iterations and the target KL divergence on the accuracy of the threshold calculation in hybrid-DE, respectively.  We compare the percentage of threshold error of hybrid-DE, with respect to  the threshold obtained from full-DE with $1000$ full-DE iterations, for  MET-LDPC codes  in Table~\ref{Table : MET_examples_codes}.  It is clear from Figs.~\ref{fig:MET_threshold_it} and~\ref{fig:MET_threshold_KL} that we can trade-off the accuracy  of threshold calculation by varying maximum number of full-DE iterations and target KL divergence accordingly.  For the purpose of comparison, variation of the percentage of threshold error of full-DE with a set of  maximum number of  full-DE iterations is also shown in the Fig.~\ref{fig:MET_threshold_it}. It is clear from Fig.~\ref{fig:MET_threshold_it}  that even when we limit the number of full-DE iterations in hybrid-DE algorithm, there is still considerable performance  improvement in terms of threshold accuracy to be gained by continuing with Gaussian approximation iterations compared to the full-DE threshold with the same maximum number of full-DE iterations but without the additional Gaussian approximation iterations.

In Figs.~\ref{Fig.MET_threshold_time}(a) and~\ref{Fig.MET_threshold_time}(b), we respectively  compare the percentage of threshold error  and CPU time gain, with respect to  the threshold and CPU time obtained from full-DE with $1000$ decoding iterations with the three single-parameter Gaussian approximation methods, and  with the hybrid-DE. We calculate the threshold using hybrid-DE method for a range of target KL divergences and  maximum number of  full-DE iterations in order to emphasize the trade-off between accuracy and efficiency.  It can be seen from Fig.~\ref{Fig.MET_threshold_time} that we can obtain up to 10 times computational  time gain by doing hybrid-DE 2 and 3, with  only loosing maximum of $5\%$  accuracy of the threshold calculation.  However, even though the all Gaussian approximation methods report a better CPU time gain than hybrid-DE, they accurately estimate the code threshold  only at higher rates, i.e., Approximation 1 and 3 estimate the code threshold with less than $5\%$ error only for code rates above 0.6 where as Approximation 2 gives an accurate estimations only at rates above 0.7.  Furthermore  it is clear from Fig.~\ref{Fig.MET_threshold_time}  that, even by doing only $10$ full-DE iterations in hybrid-DE (hybrid-DE 1), we can  still get a considerable  accuracy improvement of threshold calculation compared to the single-parameter Gaussian approximation methods. These make hybrid-DE more suitable for  code design where  accurate and efficient threshold calculation is particularly valuable.

%figure for threshold variation with iterations

%figure for threshold variation with KL divergence

%figure for threshold variation and CPU gain  with rate

%figure for code designing using GA
\begin{figure}[t]
	\centering
	\includegraphics[width=1\linewidth]{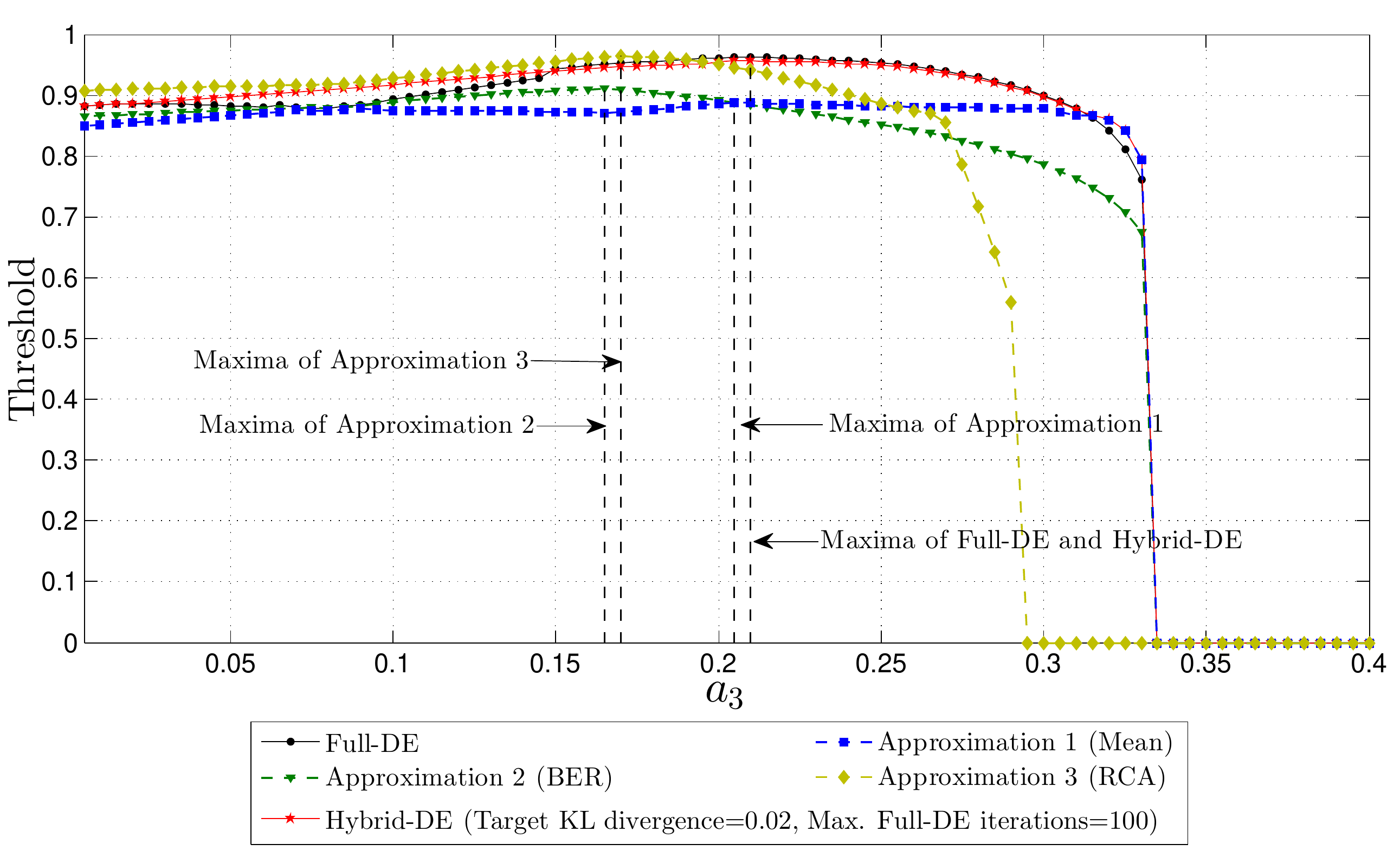}
	\vspace*{-1em}
	\caption{Threshold found through exhaustive search for a rate $1/2$ MET-LDPC code with $L(\boldsymbol{r},\boldsymbol{x})=a_1r_1x_1^2+a_2r_1x_1^3+a_3r_0x_2^3x_3^3+a_4r_1x_4$ for $a_1=0.5, a_3=a_4$ and $a_2= (1-a_1-a_4)$. }
	\label{fig:optimzation_MET}
	%\vspace{-2.5em}
\end{figure}

\subsection{Design of MET-LDPC codes} \label{MET-LDPC code design}

The aim of this section is to show how approximate DE algorithms affect the design of optimal  MET-LDPC code ensembles (defined by the  degree distribution with the largest possible code threshold). This is a non-linear cost function maximization problem, where  the cost function is the code threshold and the degree distributions are the variables to be optimized. It is still possible to obtain an optimal  degree distribution even if the DE approximation  returns an inaccurate threshold, as long as it returns the highest threshold for the optimal degree distributions. However this is not the case using Gaussian approximations. For example, we perform an exhaustive search on a single parameter ($a_3$) of a MET-LDPC code with the remaining parameters fixed in Fig.~\ref{fig:optimzation_MET}. The maxima of the  full-DE does not coincide with the maxima of the approximations. While the hybrid-DE cost function closely follows the shape of the full-DE cost function the other approximations do not. This threshold difference between  full-DE and Gaussian approximations can significantly impact the search for good code ensembles for given design constraints.

\footnotetext[6]{Threshold error = $\vline1-\frac{\sigma_{\text{App}}^*}{\sigma_{\text{DE}}^*}\vline$ where, $\sigma_{\text{App}}^*$ is the threshold calculated using relevant  method and  $\sigma_{\text{DE}}^*$ is the threshold calculated using  full-DE with the full 1000 iterations.}
\footnotetext[7]{CPU time gain = $\vline\frac{\text{CPU-time}_{\text{DE}}}{\text{CPU-time}_{\text{App}}}\vline$. Algorithms were written in Matlab and run on an Intel Xeon E5-2650,  2.6 GHz PC. The maximum number of decoding iterations were the same for all the  MET-LDPC codes considered. }

%table for rate 1/10 optimzation result
\begin{table*}[t]
	\renewcommand{\arraystretch}{1.5}
	\caption{Optimization of rate $1/10$ MET-LDPC codes on BI-AWGN channel} %\vspace{-2em}
	\label{Table : MET_examples_optimzation_rate 1/10}
	\centering
   \scriptsize
	\begin{tabular}{|c |l| c| c| }
		\hline
		\multicolumn{1}{|c|}{\multirow{2}[2]{*}{Design with}} & 	\multicolumn{1}{c|}{\multirow{2}[2]{*}{MET-LDPC code}} & \multicolumn{2}{c|}{\multirow{2}[-2]{*}{Threshold}} \\
		\cline{3-4}
		\multicolumn{1}{|c|}{\multirow{2}[2]{*}{}} & 	\multicolumn{1}{c|}{\multirow{2}[2]{*}{}} & \multicolumn{1}{c|}{\multirow{2}[-2]{*}{$\sigma_{\text{App}}^*$}} & \multicolumn{1}{c|}{\multirow{2}[-2]{*}{$\sigma_{\text{DE}}^*$ }}\\
		\hline
		\hline
		\multicolumn{1}{|c|}{\multirow{2}[-2]{*}{Reference code}} & $L(\boldsymbol{r},\boldsymbol{x}) = 0.1 r_1 x_1^3 x_2^{20} + 0.025 r_1 x_1^3 x_2^{25} + 0.875 r_1 x_3 $ & \multicolumn{1}{c|}{\multirow{2}[2]{*}{-}} & \multicolumn{1}{c|}{\multirow{2}[2]{*}{\textbf{2.5346}}}\\
		\multicolumn{1}{|c|}{\multirow{2}[-3]{*}{(Table X of~\cite{RichardsonW2002multi})}} & $R(\boldsymbol{x}) = 0.025  x_1^{15}  + 0.875  x_2^3 x_3 $ & \multicolumn{1}{c|}{\multirow{2}[2]{*}{}}  & \multicolumn{1}{c|}{\multirow{2}[2]{*}{}}\\
		\hline
		\multicolumn{1}{|c|}{\multirow{2}[2]{*}{Full-DE}} & $L(\boldsymbol{r},\boldsymbol{x}) = 0.0775 r_1 x_1 x_2 x_3^{21} + 0.0477 r_1 x_1^2 x_2 x_3^{20} + 0.8747 r_1 x_4 $ & \multicolumn{1}{c|}{\multirow{2}[2]{*}{-}} & \multicolumn{1}{c|}{\multirow{2}[2]{*}{\textbf{2.5424}}}\\
		\multicolumn{1}{|c|}{\multirow{2}[2]{*}{}} & $R(\boldsymbol{x}) = 0.0011  x_1^6 x_2^4 + 0.0028  x_1^6 x_2^5 + 0.0214  x_1^7 x_2^5 + 0.0412  x_3^2 x_4 + 0.8335  x_3^3 x_4 $ & \multicolumn{1}{c|}{\multirow{2}[2]{*}{}}  & \multicolumn{1}{c|}{\multirow{2}[2]{*}{}}\\
		\hline
		\multicolumn{1}{|c|}{\multirow{2}[2]{*}{Hybrid-DE}} & $L(\boldsymbol{r},\boldsymbol{x}) = 0.0538 r_1 x_1^3 x_2 x_3^{20} + 0.0775 r_1 x_1 x_2 x_3^{19} + 0.8687 r_1 x_4 $ & \multicolumn{1}{c|}{\multirow{2}[2]{*}{2.5455}} & \multicolumn{1}{c|}{\multirow{2}[2]{*}{\textbf{2.5372}}}\\
		\multicolumn{1}{|c|}{\multirow{2}[2]{*}{}} & $R(\boldsymbol{x}) = 0.0116  x_1^7 x_2^4 + 0.0137  x_1^8 x_2^4 + 0.0061  x_1^8 x_2^5 + 0.0573  x_3^2 x_4 + 0.8113  x_3^3 x_4 $ & \multicolumn{1}{c|}{\multirow{2}[2]{*}{}}  & \multicolumn{1}{c|}{\multirow{2}[2]{*}{}}\\
		\hline
		\multicolumn{1}{|c|}{\multirow{2}[-2]{*}{Approximation 1}} & $L(\boldsymbol{r},\boldsymbol{x}) = 0.0544 r_1 x_1^2 x_3^{20} + 0.0641 r_1 x_1^3 x_2 x_3^{25} + 0.8815 r_1 x_4 $ & \multicolumn{1}{c|}{\multirow{2}[2]{*}{2.4661}} & \multicolumn{1}{c|}{\multirow{2}[2]{*}{\textbf{2.4965}}}\\
		\multicolumn{1}{|c|}{\multirow{2}[-3]{*}{(Mean)}} & $R(\boldsymbol{x}) = 0.0099  x_1^{16} x_2^3 + 0.0035  x_1^{16} x_2^3 + 0.0051  x_1^{17} x_2^4 + 0.8355  x_3^3 x_4 + 0.0460  x_3^4 x_4 $ & \multicolumn{1}{c|}{\multirow{2}[2]{*}{}}  & \multicolumn{1}{c|}{\multirow{2}[2]{*}{}}\\
		\hline
		\multicolumn{1}{|c|}{\multirow{2}[-2]{*}{Approximation 2}} & $L(\boldsymbol{r},\boldsymbol{x}) = 0.06 r_1 x_1^2 x_2^{19} + 0.0576 r_1 x_1^3 x_2^{23}  + 0.8824 r_1 x_3 $ & \multicolumn{1}{c|}{\multirow{2}[2]{*}{2.3659}} & \multicolumn{1}{c|}{\multirow{2}[2]{*}{ \textbf{2.3850}}}\\
		\multicolumn{1}{|c|}{\multirow{2}[-3]{*}{(BER)}} & $R(\boldsymbol{x}) = 0.0058  x_1^{16} + 0.0118  x_1^{17}  + 0.1833  x_2^{2} x_3 + 0.6991  x_2^3 x_3$ & \multicolumn{1}{c|}{\multirow{2}[2]{*}{}}  & \multicolumn{1}{c|}{\multirow{2}[2]{*}{}}\\
		\hline
		\multicolumn{1}{|c|}{\multirow{2}[-2]{*}{Approximation 3}} & $L(\boldsymbol{r},\boldsymbol{x}) = 0.0942  r_1 x_1^2 x_2 x_3^{20} + 0.0336 r_1 x_1 x_2 x_3^{21} + 0.8722 r_1 x_4 $ & \multicolumn{1}{c|}{\multirow{2}[2]{*}{2.5056}} & \multicolumn{1}{c|}{\multirow{2}[2]{*}{ \textbf{2.5303}}}\\
		\multicolumn{1}{|c|}{\multirow{2}[-3]{*}{(RCA)}} & $R(\boldsymbol{x}) = 0.0006  x_1^{7} x_2^4 + 0.0107  x_1^{8} x_2^4 + 0.0165  x_1^{8} x_2^5 + 0.0262  x_3^2 x_4 + 0.8459  x_3^3 x_4$ & \multicolumn{1}{c|}{\multirow{2}[2]{*}{}}  & \multicolumn{1}{c|}{\multirow{2}[2]{*}{}}\\
		\hline		
	\end{tabular}
	\vspace{1.5em}
\end{table*}
\normalsize

%table for rate 1/2 code optimization result
\begin{table*}[t]
	\renewcommand{\arraystretch}{1.5}
	\caption{Optimization of rate $1/2$ MET-LDPC codes on BI-AWGN channel} %\vspace{-2em}
	\label{Table : MET_examples_optimzation_rate 1/2}
	\centering
\scriptsize
	\begin{tabular}{|c |l| c| c| }
		\hline
		\multicolumn{1}{|c|}{\multirow{2}[2]{*}{Design with}} & 	\multicolumn{1}{c|}{\multirow{2}[2]{*}{MET-LDPC code}} & \multicolumn{2}{c|}{\multirow{2}[-2]{*}{Threshold}} \\
		\cline{3-4}
		\multicolumn{1}{|c|}{\multirow{2}[2]{*}{}} & 	\multicolumn{1}{c|}{\multirow{2}[2]{*}{}} & \multicolumn{1}{c|}{\multirow{2}[-2]{*}{$\sigma_{\text{App}}^*$}} & \multicolumn{1}{c|}{\multirow{2}[-2]{*}{$\sigma_{\text{DE}}^*$}}\\
		\hline
		\hline
		\multicolumn{1}{|c|}{\multirow{2}[-2]{*}{Reference code}} & $L(\boldsymbol{r},\boldsymbol{x}) = 0.2 r_0 x_2^3 x_3^{3} + 0.5 r_1 x_1^2 + 0.3 r_1 x_1^3 + 0.2 r_1 x_4 $ & \multicolumn{1}{c|}{\multirow{2}[2]{*}{-}} & \multicolumn{1}{c|}{\multirow{2}[2]{*}{\textbf{0.9656}}}\\
		\multicolumn{1}{|c|}{\multirow{2}[-3]{*}{(Table VI of~\cite{RichardsonW2002multi})}} & $R(\boldsymbol{x}) = 0.1  x_1^{3} x_2^2  + 0.4  x_1^{4} x_2 + 0.2  x_3^3 x_4 $ & \multicolumn{1}{c|}{\multirow{2}[2]{*}{}} & \multicolumn{1}{c|}{\multirow{2}[2]{*}{}}\\
		\hline
		\multicolumn{1}{|c|}{\multirow{2}[2]{*}{Full-DE}} & $L(\boldsymbol{r},\boldsymbol{x}) = 0.4162 r_0 x_1 x_2^2 x_3^{2} + 0.5629 r_1 x_1^2  + 0.0294 r_1 x_1^3  +0.4076 r_1 x_4$ & \multicolumn{1}{c|}{\multirow{2}[2]{*}{-}} & \multicolumn{1}{c|}{\multirow{2}[2]{*}{\textbf{0.9713}}}\\
		\multicolumn{1}{|c|}{\multirow{2}[2]{*}{}} & $R(\boldsymbol{x}) = 0.1848 x_1^3 x_2 + 0.2191 x_1^3 x_2^2 + 0.1047 x_1^4 x_2^2 + 0.3905 x_3^2 x_4 + 0.0171 x_3^3 x_4 $ & \multicolumn{1}{c|}{\multirow{2}[2]{*}{}} & \multicolumn{1}{c|}{\multirow{2}[2]{*}{}}\\
		\hline
		\multicolumn{1}{|c|}{\multirow{2}[2]{*}{Hybrid-DE}} & $L(\boldsymbol{r},\boldsymbol{x}) = 0.5962 r_0 x_2^2 x_3^3  + 0.0004 r_1 x_1^2 x_2^3 x_3  + 0.1055 r_1 x_1^3  +0.8941 r_1 x_4$ & \multicolumn{1}{c|}{\multirow{2}[2]{*}{0.9660}} & \multicolumn{1}{c|}{\multirow{2}[2]{*}{\textbf{0.9688}}}\\
		\multicolumn{1}{|c|}{\multirow{2}[2]{*}{}} & $R(\boldsymbol{x}) = 0.0189 x_1 x_2^5 + 0.0679 x_1 x_2^6 + 0.1153 x_1^2 x_2^6 + 0.8935 x_3^2 x_4 + 0.0006 x_3^3 x_4$ & \multicolumn{1}{c|}{\multirow{2}[2]{*}{}} & \multicolumn{1}{c|}{\multirow{2}[2]{*}{}}\\
		\hline
		\multicolumn{1}{|c|}{\multirow{2}[-2]{*}{Approximation 1}} & $L(\boldsymbol{r},\boldsymbol{x}) = 0.2792 r_0 x_2^3 x_3^{3} + 0.4067 r_1 x_1^2 + 0.2341 r_1 x_1^3 + 0.3592 r_1 x_4  $ & \multicolumn{1}{c|}{\multirow{2}[2]{*}{0.9152}} & \multicolumn{1}{c|}{\multirow{2}[2]{*}{\textbf{0.9588}}}\\
		\multicolumn{1}{|c|}{\multirow{2}[-3]{*}{(Mean)}} & $R(\boldsymbol{x}) = 0.0024  x_1^{3} x_2 + 0.1618  x_1^{3} x_2^2 + 0.2558  x_1^4 x_2^2 + 0.2401 x_3^2 x_4 + 0.1191  x_3^3 x_4$ & \multicolumn{1}{c|}{\multirow{2}[2]{*}{}} & \multicolumn{1}{c|}{\multirow{2}[2]{*}{}}\\
		\hline
		\multicolumn{1}{|c|}{\multirow{2}[-2]{*}{Approximation 2}} & $L(\boldsymbol{r},\boldsymbol{x}) =  0.5034 r_0 x_1 x_2 x_3^{3} + 0.0068 r_1 x_1^2 x_2 x_3  + 0.2337 r_1 x_1^3 + 0.7595 r_1 x_4   $ & \multicolumn{1}{c|}{\multirow{2}[2]{*}{ 0.9099}} & \multicolumn{1}{c|}{\multirow{2}[2]{*}{\textbf{0.9535}}}\\
		\multicolumn{1}{|c|}{\multirow{2}[-3]{*}{(BER)}} & $R(\boldsymbol{x}) = 0.0016  x_1^{4} x_2^2  + 0.2201  x_1^{5} x_2^2 + 0.0223  x_1^5 x_2^3 + 0.0018 x_3 x_4 + 0.7576  x_3^2 x_4$ & \multicolumn{1}{c|}{\multirow{2}[2]{*}{}} & \multicolumn{1}{c|}{\multirow{2}[2]{*}{}}\\
		\hline
		\multicolumn{1}{|c|}{\multirow{2}[-2]{*}{Approximation 3}} & $L(\boldsymbol{r},\boldsymbol{x}) = 0.1564 r_0 x_2^3 x_3^{3} + 0.3689 r_1 x_1^2  + 0.4607 r_1 x_1^3 + 0.1704 r_1 x_4 $ & \multicolumn{1}{c|}{\multirow{2}[2]{*}{0.9435}} & \multicolumn{1}{c|}{\multirow{2}[2]{*}{\textbf{0.9420}}}\\
		\multicolumn{1}{|c|}{\multirow{2}[-3]{*}{(RCA)}} & $R(\boldsymbol{x}) = 0.0168  x_1^{4}   + 0.2934  x_1^{4} x_2 + 0.1758  x_1^5 x_2 + 0.0419 x_3^2 x_4 + 0.1285  x_3^3 x_4$ & \multicolumn{1}{c|}{\multirow{2}[2]{*}{}} & \multicolumn{1}{c|}{\multirow{2}[2]{*}{}}\\
		\hline		
	\end{tabular}
%	\vspace{1em}
\end{table*}
\normalsize

To further demonstrate the effect of the Gaussian approximations on code design, we  design rate $1/10$ and $1/2$ MET-LDPC codes on BI-AWGN channel with  full-DE, hybrid-DE and the three Gaussian approximations stated in Section~\ref{Gaussian app}. We use the joint optimization methodology proposed in~\cite{sachiniITW2014optimization} to design MET-LDPC codes. The results are presented in Tables~\ref{Table : MET_examples_optimzation_rate 1/10} and~\ref{Table : MET_examples_optimzation_rate 1/2} where the values are rounded off to four decimal places. For a fair comparison, we consider  similar MET-LDPC code structures from  Table X and VI of~\cite{RichardsonW2002multi} for rate $1/10$  and $1/2$ MET-LDPC codes respectively. The maximum number of decoding iterations and target bit error rate for the BP decoding process is set to $1000$ and $10^{-10}$ respectively.  For hybrid-DE,  we set the target KL divergence to $0.04$ and  the maximum number of  full-DE iterations allowed to $100$ and calculate KL divergence  after every 5 decoding iterations to check whether the message PDFs are close to Gaussian. The results in Table~\ref{Table : MET_examples_optimzation_rate 1/10} show that  Approximations 1 and  2 can result in noticeable inaccuracy for designing  rate $1/10$ MET-LDPC codes.  However, the  rate $1/10$  MET-LDPC code designed using  hybrid-DE closely matches  the MET-LDPC code designed with  full-DE.  The results in Table~\ref{Table : MET_examples_optimzation_rate 1/2} show that   the Approximations 1 and 2  are more  successful at designing  rate $1/2$ MET-LDPC codes and the worst performing algorithm in this case was Approximation 3. Nevertheless, the rate $1/2$ MET-LDPC code designed with  hybrid-DE gives the closest match to the  MET-LDPC code designed with  full-DE.

\begin{figure}[t]
	\centering
	\includegraphics[width=1\linewidth]{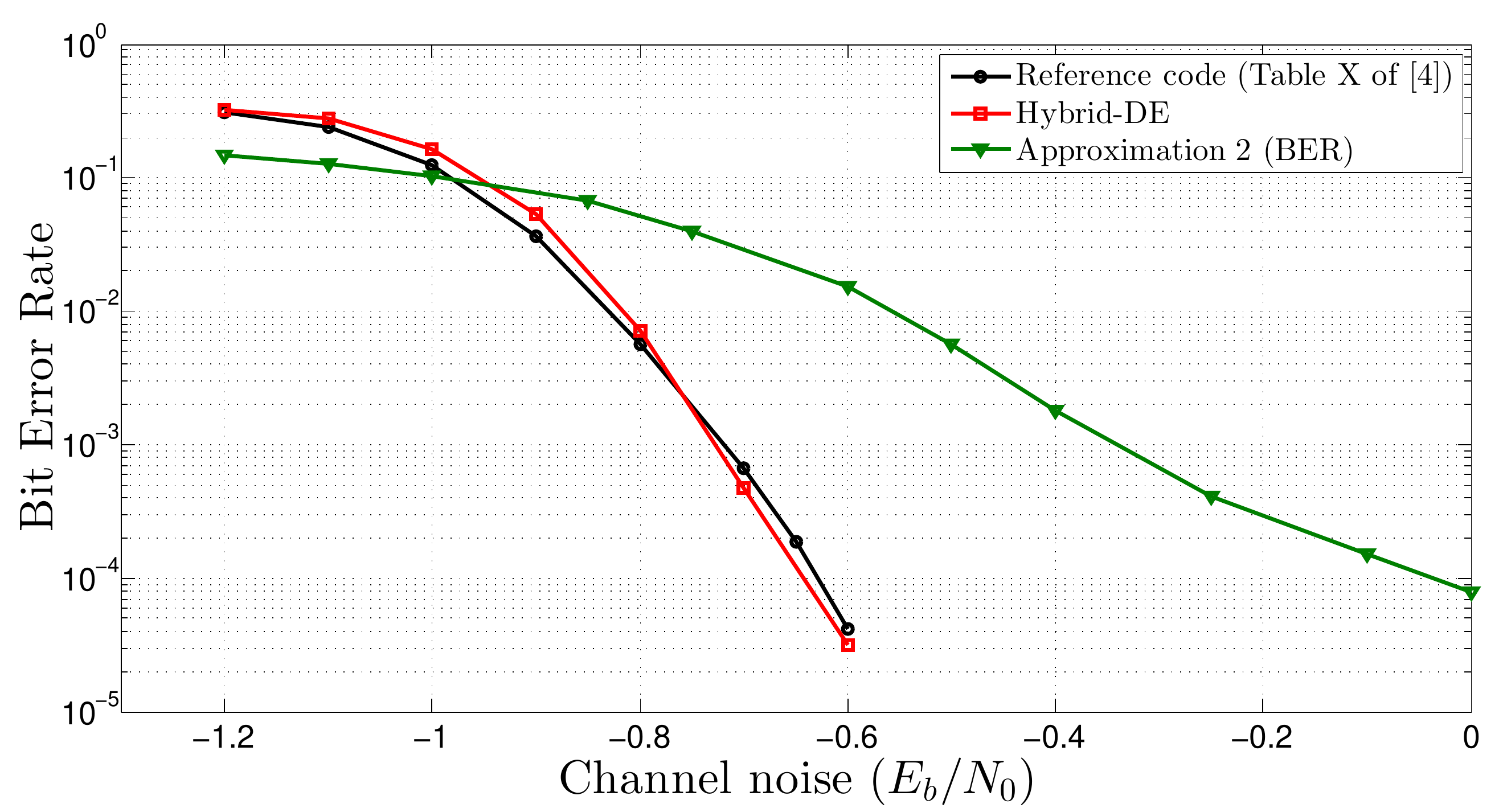}
	\vspace*{-1em}
	\caption{The bit error rate performance of length-$100000$,  rate $1/10$ MET-LDPC codes with degree distributions from Table~\ref{Table : MET_examples_optimzation_rate 1/10} on BI-AWGN channel. }
	\label{fig:BER}
\end{figure}

We then simulate the finite-length performances for the rate $1/10$ MET-LDPC codes with degree distributions from Table~\ref{Table : MET_examples_optimzation_rate 1/10} with block length of $100000$. As expected, the threshold differences between the ensembles shown in Table~\ref{Table : MET_examples_optimzation_rate 1/10} are  reflected in the finite-length performance differences in Fig.~\ref{fig:BER}.

This suggests that our proposed hybrid-DE method performs similarly to  full-DE,  making it suitable for  code optimization even at low rate and with punctured variable nodes. Since our proposed method can also be used to  strike a balance between  efficiency and  accuracy required, we  claim that the proposed hybrid method is a suitable DE approximation technique for code design.

%***********************************************************************************************
%***************Section VII - Conclusion********************************************************
%***********************************************************************************************

\section{Conclusion} \label{Conclusion}
This paper investigated the performance of  density evolution for low-density parity-check (LDPC) and multi-edge type low-density parity-check (MET-LDPC)  codes over the  binary input additive white Gaussian noise  channel. We applied and analyzed three single-parameter  Gaussian approximation models. We showed that the accuracy of single-parameter Gaussian approximations might be poor under several conditions, namely codes at low rates and codes with punctured variable nodes.  Then, we proposed a more accurate density evolution (DE) approximation, referred to as  hybrid-DE, which is a combination of the  full-DE and a single-parameter Gaussian approximation. With hybrid-DE, we avoided the symmetric Gaussian assumption at early decoding iterations of BP decoding, making our code threshold calculations  significantly more accurate than existing methods of using Gaussian approximations for all decoding iterations. At the same time, hybrid-DE significantly reduced the computational time of evaluating the code threshold compared to  full-DE.  These make hybrid-DE more suitable for the code design where  accurate and efficient threshold calculation is particularly valuable. Finally, we considered code optimization and presented a code design by using  full-DE, hybrid-DE and three Gaussian approximations. The designed codes using hybrid-DE closely match with the codes designed using  full-DE. Thus, we can suggest that the hybrid-DE is a good DE technique for code design. Since hybrid-DE is not specific to MET-LDPC codes, it also can be used for designing other codes defined on graphs such as irregular LDPC codes.

\newpage

\vspace*{0.5em}
\appendix[] \label{appendix A} %\mbox{}
\vspace*{-0.5em}
\begin{table}[!h]
	\renewcommand{\arraystretch}{1.2}
	\caption{MET-LPDC codes used in Figs.~\ref{fig:MET_threshold_it} to~\ref{Fig.MET_threshold_time}} %\vspace{-2em}
	\label{Table : MET_examples_codes}
	\centering
	\scriptsize
	%\begin{tabular}{|C{1cm} |C{0.6cm}| C{1cm}| C{0.9cm} |C{0.7cm} |C{0.9cm} |C{0.7cm}| C{0.9cm}| C{0.7cm}| C{0.9cm} |C{0.7cm} |C{0.9cm}| C{0.7cm}| }
	\begin{tabular}{|c |c| l|  }
		\hline
		Code & Rate & \multicolumn{1}{c|}{Degree distribution} \\
		\hline
		\hline
		\multicolumn{1}{|c|}{\multirow{4}[2]{*}{Code }} & \multicolumn{1}{c|}{\multirow{4}[2]{*}{0.1}}  & $L(\boldsymbol{r},\boldsymbol{x}) =$  \\
		\multicolumn{1}{|c|}{\multirow{4}[2]{*}{A}} & \multicolumn{1}{c|}{\multirow{4}[2]{*}{}} & $0.6737r_1x_1^2 + 0.3263r_1x_1^3+ 0.0001r_0x_2^3x_3^3 + 0.0001r_1x_4$ \\						
		\multicolumn{1}{|c|}{\multirow{4}[2]{*}{}} & \multicolumn{1}{c|}{\multirow{4}[2]{*}{}} & $R(\boldsymbol{x}) = $ \\
		\multicolumn{1}{|c|}{\multirow{4}[2]{*}{}} & \multicolumn{1}{c|}{\multirow{4}[2]{*}{}} & $ 0.3737x_1^2 + 0.5260x_1^3 + 0.0003x_1^3x_2 +  0.0001x_3^3x_4$ \\
		\hline
		
		\multicolumn{1}{|c|}{\multirow{4}[2]{*}{Code }} & \multicolumn{1}{c|}{\multirow{4}[2]{*}{0.2}}  &$L(\boldsymbol{r},\boldsymbol{x}) = $  \\
		\multicolumn{1}{|c|}{\multirow{4}[2]{*}{B}} & \multicolumn{1}{c|}{\multirow{4}[2]{*}{}} & $0.7281r_1x_1^2 + 0.0052r_1x_1^3+ 0.2669 r_0x_2^3x_3^3 + 0.2669r_1x_4$ \\
		\multicolumn{1}{|c|}{\multirow{4}[2]{*}{}} & \multicolumn{1}{c|}{\multirow{4}[2]{*}{}} & $R(\boldsymbol{x}) = $ \\
		\multicolumn{1}{|c|}{\multirow{4}[2]{*}{}} & \multicolumn{1}{c|}{\multirow{4}[2]{*}{}} & $ 0.1284x_1x_2 + 0.6711x_1^2x_2 + 0.0006x_1^2x_2^2 +  0.2669x_3^3x_4$ \\
		\hline
		
		\multicolumn{1}{|c|}{\multirow{4}[2]{*}{Code }} &  \multicolumn{1}{c|}{\multirow{4}[2]{*}{0.3}}  & $L(\boldsymbol{r},\boldsymbol{x}) = $ \\
		\multicolumn{1}{|c|}{\multirow{4}[2]{*}{C}} & \multicolumn{1}{c|}{\multirow{4}[2]{*}{}} & $0.7213r_1x_1^2 + 0.0006r_1x_1^3+ 0.2781r_0x_2^3x_3^3 + 0.2781r_1x_4$ \\
		\multicolumn{1}{|c|}{\multirow{4}[2]{*}{}} & \multicolumn{1}{c|}{\multirow{4}[2]{*}{}} & $R(\boldsymbol{x}) = $ \\
		\multicolumn{1}{|c|}{\multirow{4}[2]{*}{}} & \multicolumn{1}{c|}{\multirow{4}[2]{*}{}} & $ 0.5656x_1^2x_2 +0.09x_1^2x_2^2 + 0.0444x_1^3x_2^2 +  0.2781x_3^3x_4$ \\
		\hline
		
		\multicolumn{1}{|c|}{\multirow{4}[2]{*}{Code }} & \multicolumn{1}{c|}{\multirow{4}[2]{*}{0.4}}  &  $L(\boldsymbol{r},\boldsymbol{x}) = $\\
		\multicolumn{1}{|c|}{\multirow{4}[2]{*}{D}} & \multicolumn{1}{c|}{\multirow{4}[2]{*}{}} & $0.6864r_1x_1^2 + 0.0007r_1x_1^3+ 0.3129 r_0x_2^3x_3^3 + 0.3129r_1x_4$ \\
		\multicolumn{1}{|c|}{\multirow{4}[2]{*}{}} & \multicolumn{1}{c|}{\multirow{4}[2]{*}{}} & $R(\boldsymbol{x}) =$ \\
		\multicolumn{1}{|c|}{\multirow{4}[2]{*}{}} & \multicolumn{1}{c|}{\multirow{4}[2]{*}{}} & $ 0.2613x_1^2x_2 + 0.1638x_1^2x_2^2+ 0.1749x_1^3x_2^2 +  0.3129x_3^3x_4$ \\
		\hline
		
		\multicolumn{1}{|c|}{\multirow{4}[2]{*}{Code }} & \multicolumn{1}{c|}{\multirow{4}[2]{*}{0.5}}  & $L(\boldsymbol{r},\boldsymbol{x}) = $ \\
		\multicolumn{1}{|c|}{\multirow{4}[2]{*}{E}} & \multicolumn{1}{c|}{\multirow{4}[2]{*}{}} &  $0.5713r_1x_1^2 + 0.1788r_1x_1^3 + 0.2497r_0x_2^3x_3^3 + 0.2497 r_1x_4$\\
		\multicolumn{1}{|c|}{\multirow{4}[2]{*}{}} & \multicolumn{1}{c|}{\multirow{4}[2]{*}{}} &  $R(\boldsymbol{x}) =$\\
		\multicolumn{1}{|c|}{\multirow{4}[2]{*}{}} & \multicolumn{1}{c|}{\multirow{4}[2]{*}{}} &  $ 0.2507x_1^3x_2 + 0.0699x_1^3x_2^2 + 0.1793x_1^4x_2^2  +   0.2497x_3^3x_4$\\
		\hline
		
		\multicolumn{1}{|c|}{\multirow{4}[2]{*}{Code }} & \multicolumn{1}{c|}{\multirow{4}[2]{*}{0.6}}  &  $L(\boldsymbol{r},\boldsymbol{x}) = $ \\
		\multicolumn{1}{|c|}{\multirow{4}[2]{*}{F}} & \multicolumn{1}{c|}{\multirow{4}[2]{*}{}} & $0.5001r_1x_1^2 + 0.3r_1x_1^3+ 0.1999 r_0x_2^3x_3^3 + 0.1999r_1x_4$ \\
		\multicolumn{1}{|c|}{\multirow{4}[2]{*}{}} & \multicolumn{1}{c|}{\multirow{4}[2]{*}{}} & $R(\boldsymbol{x}) =$ \\
		\multicolumn{1}{|c|}{\multirow{4}[2]{*}{}} & \multicolumn{1}{c|}{\multirow{4}[2]{*}{}} & $ 0.0998x_1^4x_2 + 0.1005x_1^5x_2^2+ 0.1997x_1^5x_2^2 +  0.1999x_3^3x_4$ \\
		\hline
		
		\multicolumn{1}{|c|}{\multirow{4}[2]{*}{Code }} & \multicolumn{1}{c|}{\multirow{4}[2]{*}{0.7}}  &  $L(\boldsymbol{r},\boldsymbol{x}) = $\\
		\multicolumn{1}{|c|}{\multirow{4}[2]{*}{G}} & \multicolumn{1}{c|}{\multirow{4}[2]{*}{}} & $0.3501r_1x_1^2 + 0.6190r_1x_1^3 + 0.0309r_0x_2^3x_3^3 + 0.0309 r_1x_4$ \\
		\multicolumn{1}{|c|}{\multirow{4}[2]{*}{}} & \multicolumn{1}{c|}{\multirow{4}[2]{*}{}} & $R(\boldsymbol{x}) = $ \\
		\multicolumn{1}{|c|}{\multirow{4}[2]{*}{}} & \multicolumn{1}{c|}{\multirow{4}[2]{*}{}} & $ 0.1428x_1^8 + 0.0645x_1^9 + 0.0927x_1^9x_2 +  0.0309x_3^3x_4$ \\
		\hline		
	\end{tabular}
	%\vspace{-2em}
\end{table}

\normalsize	

\footnotesize
\singlespacing	
\bibliographystyle{IEEEtran}
%\bibliography{IEEEabrv,Bibliography}

\end{document}